\documentclass[review]{elsarticle}

\usepackage{amssymb}
\usepackage{amsthm}
\usepackage{bigints}
\usepackage{lineno}
\usepackage{hyperref}
\usepackage{algorithm}
\usepackage{algorithmic}

\usepackage{amsmath}
\usepackage{xcolor}
\usepackage{multirow}
\usepackage{soul}
\usepackage{bm}
\usepackage[mathscr]{euscript}

\usepackage{makecell}
\usepackage{siunitx}

\journal{Computer Methods in Applied Mechanics and Engineering}

\begin{document}

\begin{frontmatter}



\newtheorem{theorem}{Theorem}[section]
\newtheorem{lemma}[theorem]{Lemma}
\newtheorem{proposition}[theorem]{Proposition}
\newtheorem{corollary}[theorem]{Corollary}
\newtheorem{definition}[theorem]{Definition}
\newtheorem{problem}[theorem]{Problem}

\title{Quadrilateral Mesh Generation III : Optimizing Singularity Configuration Based on Abel-Jacobi Theory}

\author[addr1,addr4]{Xiaopeng Zheng}
\author[addr1]{Yiming Zhu}
\author[addr1,addr2]{Na Lei\corref{correspondingauthor}}
\cortext[correspondingauthor]{Corresponding author}
\ead{nalei@dlut.edu.cn}
\author[addr1,addr4]{Zhongxuan Luo}
\author[addr3]{Xianfeng Gu}

\address[addr1]{Dalian University of Technology, Dalian, China}
\address[addr3]{Stony Brook University, New York, US}
\address[addr4]{Key Laboratory for Ubiquitous Network and Service Software of Liaoning Province, Dalian, China}
\address[addr2]{DUT-RU Co-Research Center of Advanced ICT for Active Life, Dalian, China}

\begin{abstract}

This work proposes a rigorous and practical algorithm for generating meromorphic quartic differentials for the purpose of quad-mesh generation. The work is based on the Abel-Jacobi theory of algebraic curve.

The algorithm pipeline can be summarized as follows: calculate the homology group; compute the holomorphic differential group;  construct the period matrix of the surface and Jacobi variety; calculate the Abel-Jacobi map for a given divisor;  optimize the divisor to satisfy the Abel-Jacobi condition by an integer programming; compute the flat Riemannian metric with cone singularities at the divisor by Ricci flow; isometric immerse the surface punctured at the divisor onto the complex plane and pull back the canonical holomorphic differential to the surface to obtain the meromorphic quartic differential; construct the motor-graph to generate the resulting T-Mesh.

The proposed method is rigorous and practical. The T-mesh results can be applied for constructing T-Spline directly. The efficiency and efficacy of the proposed algorithm are demonstrated by experimental results.

\end{abstract}


\end{frontmatter}


\section{Introduction}

In computational mechanics, Computer Aided Design, geometric design, computer graphics, medical imaging, digital geometry processing and many other engineering fields, quadrilateral mesh is a universal and crucial boundary surface representation. Although quadrilateral meshes have been broadly applied in the real industrial world, the theoretic understanding of their geometric structures remains primitive. 
Recently, \cite{} makes a breakthrough 
from Algebraic geometric view, basically a quad-mesh induces a conformal structure and can be treated as a Riemann surface. Furthermore, a quad-mesh is equivalent to a meromorphic quartic differential with closed trajectories, and the singularities satisfy the Abel-Jacobi condition. This discovery provides a solid theoretic foundation for quad-meshing.

\subsection{Abel-Jacobi Condition} 

Suppose a surface $(\Sigma,\mathbf{g})$ is embedded in the Euclidean space $\mathbb{R}^3$ with the induced Euclidean Riemannian metric $\mathbf{g}$. Suppose the surface is represented as a quadrilateral mesh $\mathcal{Q}$, then $\mathcal{Q}$ induces a special combinatorial structure, a Riemannian metric structure, and a conformal structure.\\
\noindent{\emph{Combinatorial structure}}: Suppose the number of vertices, edges, faces of $\mathcal{Q}$ are $V,E,F$, then $E=2F$ and the Euler formula holds, $V+F-E=\chi(\Sigma)$, where $\chi(\Sigma)$ is the Euler characteristic number of $\Sigma$. The vertices with topological valence $4$ are called normal, otherwise singular.\\
\noindent{\emph{Riemannian metric structure}}: A flat metric with cone singularities $\mathbf{g}_Q$ can be induced by $\mathcal{Q}$ by treating each face as a unit planar square. A vertex with $k$-valence has the discrete curvature $(4-k)/2\pi$, the total curvature satisfies the \emph{Guass-Bonnet condition}:
\begin{equation}
    \sum_{v} \frac{4-\text{val}(v)}{2} \pi = 2\pi\chi(\Sigma),
    \label{eqn:gauss_bonnet}
\end{equation}
where $\text{val}(v)$ is the topological valence of $v$.
The holonomy group induced by the metric $\mathbf{g}_Q$ on the surface $\Sigma\setminus \mathcal{S}$ with punctures at the singular vertices $\mathcal{S}$ is the rotation group 
\begin{equation}
\text{Hol}(\Sigma\setminus \mathcal{S},\mathbf{g}_Q)=\{e^{i\frac{\pi}{2}k},k\in \mathbb{Z}\}.
\label{eqn:holonomy}
\end{equation}
This is the so-called \emph{holonomy condition} \cite{CMAME_Quad_Mesh_I}.\\
\noindent{\emph{Conformal structure}}: The quad-mesh
$\mathcal{Q}$ induces a conformal structure, and can be treated as a Riemann surface $S_Q$; furthermore, it induces a meromorphic quartic differential $\omega_Q$, whose horizontal and vertical trajectories are finite. The valence-3 and valence-5 singularities of $\mathcal{Q}$ are the poles and zeros of $\omega_Q$, and the divisor of $\omega_Q$ represents the configuration of singularities of $\mathcal{Q}$, denoted as $(\omega_Q)$. Suppose $\varphi$ is a holomorphic 1-form on $S_Q$, then $\varphi^4$ is a holomorphic quartic differential, then $(\omega_Q)$ and $4(\varphi)$ are equivalent, and satisfy the Abel-Jacobi condition, the image of the Abel-Jacobi map is zero in the Jacobian variety $(J(S_Q))$, $\mu((\omega_Q)-4(\varphi))=0$.

\subsection{Construct Meromorphic Quartic Differential}
The procedure to generate quadrilateral meshes can be summarized as follows: 1) feature points location, and the features are used as part of the singularities of $\mathcal{Q}$; 2) improve the initial singularity set to satisfy the Abel-Jacobi condition to obatin $\mathcal{S}$; 3) construct a meromorphic quadratic differential $\omega$, whose divisor $(\omega)$ equals to $\mathcal{S}$; 4) deform the conformal structure such that the horizontal and vertical trajectories of $\omega$ are closed; 5) trace the horizontal and vertical trajectories of $\omega$ to form the quad-mesh $\mathcal{Q}$.

This work focuses on the first 3 steps. If the initial divisor doesn't satisfies the Gauss-Bonnet condition \ref{eqn:gauss_bonnet}, we will add more poles and zeros at the critical points of Gaussian curvature. Then we minimize the squared norm of the Abel-Jacobi map image of the divisor using gradient descend algorithm. Once the divisor satisfies the Abel-Jacobi condition, we use surface Ricci flow to compute a flat cone metric which concentrates all the curvature at the poles and zeros of the divisor. We isometrically immerse the surface punctured at the divisor into the plane, and pull the canonical holomorphic quartic differential $(dz)^4$ on $\mathbb{C}$ back to the surface, to get the desired meromorphic quartic differential. The meromorphic quartic differential can be applied to generate T-mesh and construct T-Splines.

\subsection{Contributions}

Based on the Abel-Jacobi theory for quad-mesh generation, this work proposes a novel algorithm to generate meromorphic quartic differential, the algorithm has solid theoretic foundation, and practically effective. Conventional methods are heuristic and involves human intervention. In contrast, the proposed method is rigorous automatic. To the best of our knowledge, this is the first work that is based on Riemann surface to construct meromorphic quartic differential for quad-meshing.

The work is organized as follows: section \ref{sec:previous_works} briefly review the most related works; section \ref{sec:theory} introduces the theoretic background; section \ref{sec:algorithm} explains the algorithm in details; the experimental results are reported in section \ref{sec:experiments}; finally, the work concludes in section \ref{sec:conclusion}.
\section{Previous works}
\label{sec:previous_works}

This section briefly review the most related works,  we refer readers to \cite{survey:Bommes2013Quad} for more thorough reviews. Quad-mesh generation has vast literature, the following we only discuss some popular approaches.

\paragraph{Triangle Mesh Conversion} Catmull-Clark subdivision method is applied to converting triangular meshes to quad-meshes, then the original vertices become singularities. Another intuitive way is to merge two triangular faces adjacent to the same edge to a quadrilateral, such proposed in~\cite{Remacle2012Blossom,Marco2010Practical,Gurung2011SQuad,Velho20014}. These type of methods can only produce unstructured quad-meshes, without much quality control.

\paragraph{Patch-Based Approach}
In order to generate semi-regular quad-meshes, this type of methods calculate the skeleton first, then partition the mesh input several quadrilateral patches, each patch is regularly tessellated into quads. There are different strategies to cluster the faces to form each patch, one way is to merge neighboring triangle faces based on the similarity among the normals, the other is based on the distance among the centers of the faces \cite{Boier2004Parameterization,Carr2006Rectangular}. Poly-cube map is a normal based method to deform the surface to a poly-cube shape, such as \cite{Xia2011Editable,Wang2008User,Lin2008Automatic,He2009A}. The Morse-Sample complex of eigenfunction of the Laplace operator naturally produce skeleton structure, which is utilized to generate quad-meshes. The spectral surface quadrangulation method applies this method in \cite{Dong2006Spectral,Huang2008Spectral}.

\paragraph{Parameterization Based Approach }
Parameteization method maps the surface onto planar domains, and construct a quad-mesh on the parameter domain, then pull back to the surface. There are different ways to compute the parameterization, such as using discrete harmonic forms \cite{Tong2006Designing}, periodic global parameterization \cite{Alliez2006Periodic} and
branched coverings method \cite{K2010QuadCover}. All these methods rely on solving elliptic partial differential equations on the surface.

\paragraph{Voronoi Based Approach} This approach put samples on the input surface, then compute Voronoi diagram on the surface using different distances. For example, if $L^p$ norm is applied, then the cells are similar to rectangles \cite{L2010Lp}. This method can only generate non-structured quad-mesh.

\paragraph{Cross field Based Approach}
This approach generate the cross field first, then by tracing the stream lines of the cross field \cite{RS:RPS:2014} or parameterization induced by the field \cite{survey:Bommes2013Quad}, the quad-mesh can be constructed. The cross fields are represented in different ways, such as N-RoSy representation\cite{Palacios2007Rotational}, period jump technique\cite{Li2006Representing} and complex value representation\cite{Kowalski2013A}. Then by minimizing the discrete analogy to the harmonic energy \cite{JFH}, the cross field can be smooth out. The work in \cite{landau} relates the Ginzberg-Landau theory with the cross field for genus zero surface case. This type of method is difficult to control the positions of the singularities and the global structure of the quad layout. Cross fields can be treated as the horizontal and vertical directions of a meromorphic quartic differential without magnitudes.

Comparing to the existing approaches, our method has explicit theoretic analysis for the singularities, the dimension of solution space. Therefore the theoretic rigor greatly improves the efficiency and efficacy for quad-mesh generation. 
\section{Theoretic Background}
\label{sec:theory}

This section briefly introduces the most related fundamental concepts and theorems.

\subsection{Basic Concepts of Riemann Surface}

\begin{definition}[Riemann Surface] Suppose $S$ is a two dimensional topological manifold, equipped with an atlas $\mathcal{A}=\{(U_\alpha,\varphi_\alpha)\}$, every local chart are complex coordinates $\varphi_\alpha:U_\alpha\to \mathbb{C}$, denoted as $z_\alpha$, and every transition map is biholomorphic,
\[
    \varphi_{\alpha\beta}:\varphi_\alpha(U_\alpha\cap U_\beta)\to \varphi_\beta(U_\alpha\cap U_\beta), \quad z_\alpha \mapsto z_\beta,
\]
then the atlas is called a conformal atlas. A topological surface with a conformal atlas is called a Riemann surface.
\end{definition}

Suppose $(\Sigma,\mathbf{g})$ is an oriented surface with a Riemannian metric $\mathbf{g}$. For each point $p\in \Sigma$, we can find a neighborhood $U(p)$, inside $U(p)$ the \emph{isothermal coordinates} $(u,v)$ can be constructed, such that $\mathbf{g}=e^{2\lambda(u,v)}(du^2+dv^2)$. The atlas formed by all the isothermal coordinates is a conformal atlas, therefore we obtain the following:
\begin{theorem}
All oriented, metric surfaces are Riemann surfaces.
\end{theorem}

\begin{definition}[Meromorphic Function on Riemann Surface] Suppose a Riemann surface $(S,\{(U_\alpha,\varphi_\alpha)\})$ is given. A complex function is defined on the surface $f:S\to \mathbb{C}\cup \{\infty\}$. If on each local chart $(U_\alpha,\varphi_\alpha)$, the local representation of the functions $f\circ \varphi_\alpha^{-1}:\mathbb{C}\to \mathbb{C}\cup\{\infty\}$ is meromorphic, then $f$ is called a meromorphic function defined on $S$.
\end{definition}
A memromorphic function can be treated as a holomorphic map from the Riemann surface to the unit sphere.
\begin{definition}[Zeros and Poles]
Given a meromorphic function $f(z)$, if its Laurent series has the form
\[
    f(z) = \sum_{n=k}^\infty a_n(z-z_0)^n,
\]
if $k>0$, then $z_0$ is called a zero point of order $k$; if $k<0$, then $z_0$ is called a pole of order $k$; if $k=0$, then $z_0$ is called a regular point. We denote $\nu_p(f)=k$.
\end{definition}

\begin{definition}[Meromorphic Differential] Given a Riemann surface $(S,\{z_\alpha\})$, $\omega$ is a meromorphic differential of order $n$, if it has local representation,
\[
    \omega = f_\alpha(z_\alpha) (dz_\alpha)^n,
\]
where $f_\alpha(z_\alpha)$ is a meromorphic function, $n$ is an integer; if $f_\alpha(z_\alpha)$ is a holomorphic function, then $\omega$ is called a holomorphic differential of order $n$. If $z_\alpha$ is a pole (or a zero) of $f_\alpha$ with order $k$, then $z_\alpha$ is called a pole (or a zero) of the meromorphic differential $\omega$ of order $k$.
\end{definition}
A holomorphic differential of order $2$ is called a \emph{holomrphic quadratic differential}; A meromorphic differential of order $4$ is called a \emph{meromorphic quartic differential}.

\begin{definition}[Divisor] The Abelian group freely generated by points on a Riemann surface is called the divisor group, every element is called a divisor, which has the form
\[
    D = \sum_p n_p p.
\]
The degree of a divisor is defined as $deg(D)=\sum_p n_p$. Suppose $D_1 = \sum_p n_p p$, $D_2 = \sum_p m_p p$, then $D_1\pm D_2 = \sum_p(n_p\pm m_p)p$; $D_1\le D_2$ if and only if for all $p$, $n_p \le m_p$.
\end{definition}

\begin{definition}[Meromorphic Differential Divisor] Suppose $\omega$ is a meromorphic differential on a Riemann surface $S$, suppose $p\in S$ is a point on $S$, we define the order of $\omega$ at $p$ as
\[
    \text{ord}_p(\omega) = \text{ord}_p(f_p),
\]
where $f_p$ is the local representation of $\omega$ in a neighborhood of $p$, $\omega= f_p(z)(dz)^n$. The divisor of $\omega$ is defined as
\[
    (\omega) = \sum_p \text{ord}_p(\omega) p.
\]
\end{definition}


\subsection{Abel-Jacobian Theorem}

\begin{figure}[h!]
\centering
\begin{tabular}{cc}
\includegraphics[width=0.67\textwidth]{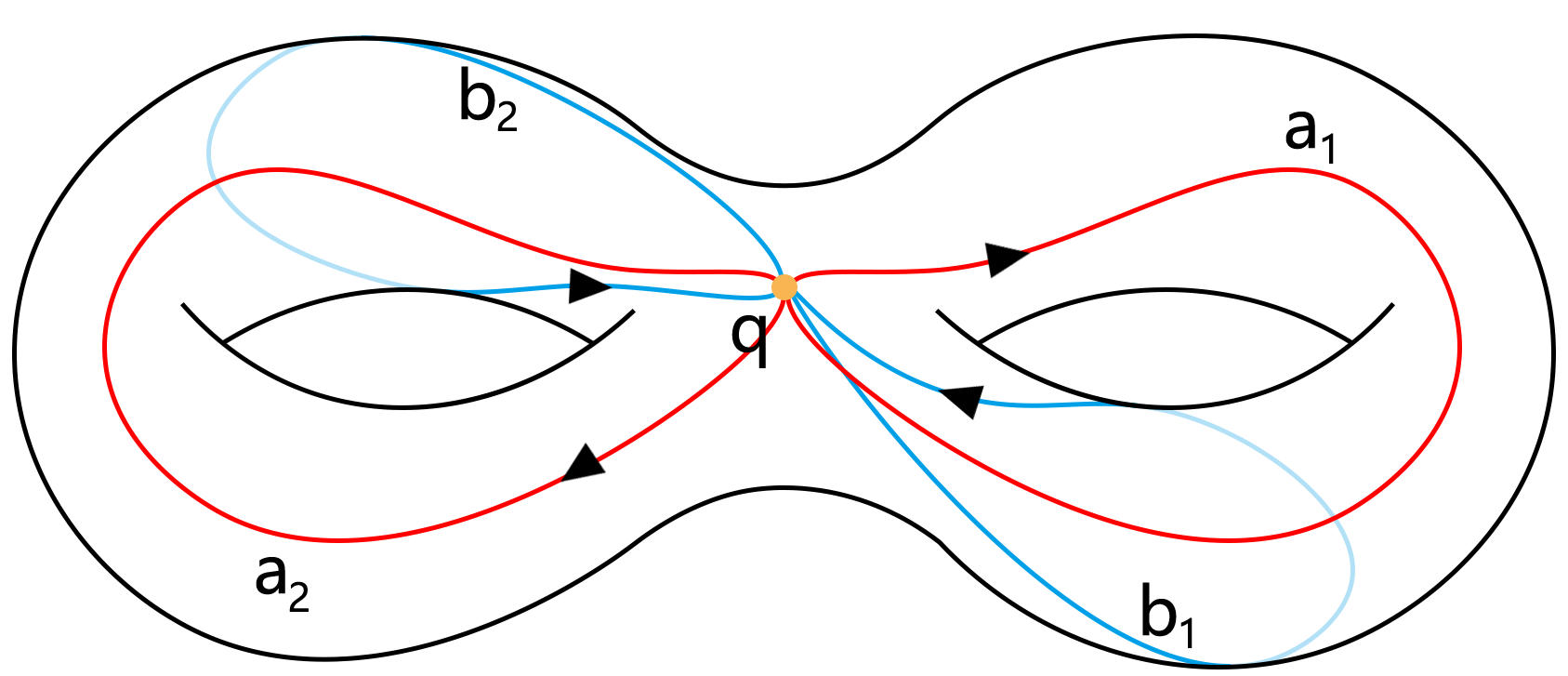}&
\includegraphics[width=0.33\textwidth]{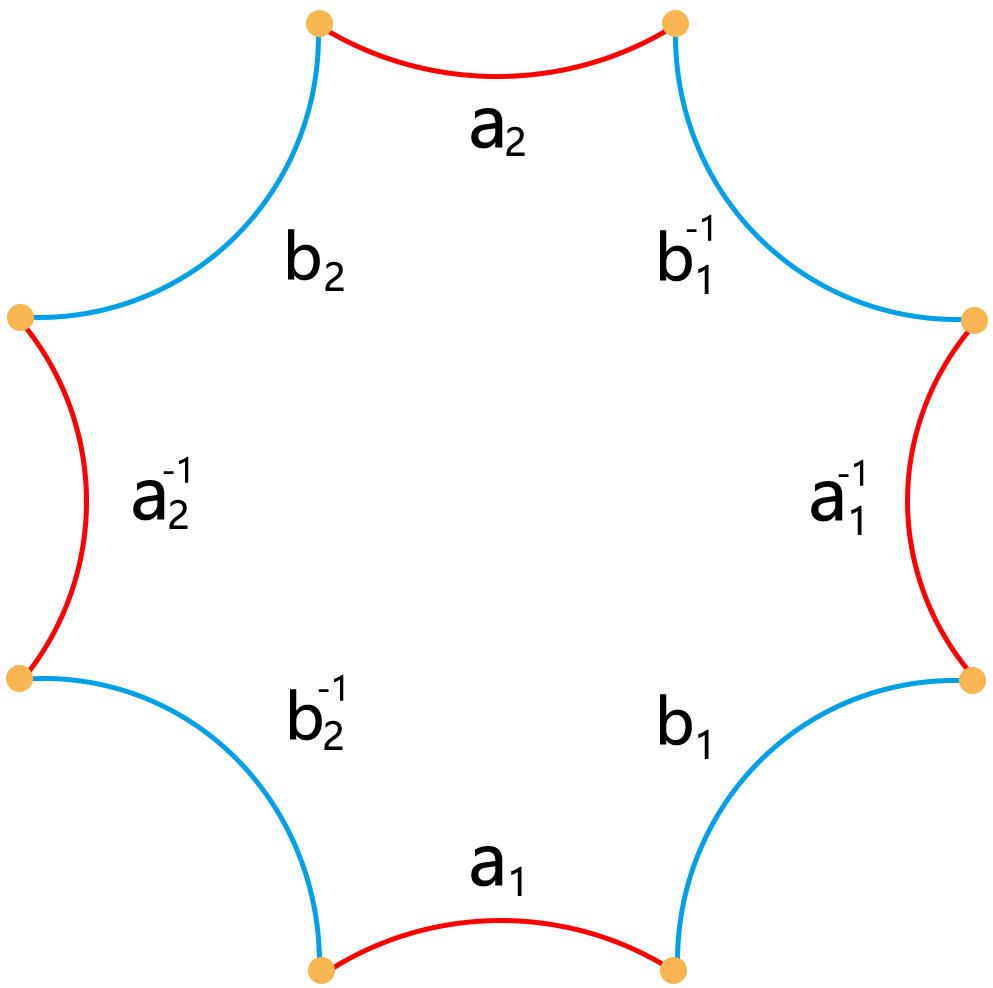}\\
\end{tabular}
\caption{Canonical fundamental group basis.}
\label{fig:fundamental group basis}
\end{figure}

Suppose $\{a_1,b_1,\dots,a_g,b_g\}$ is  a set of canonical basis for the homology group $H_1(S,\mathbb{Z})$ as shown in Fig.~\ref{fig:fundamental group basis}. Each $a_i$ and $b_i$ represent the curves around the inner and outer circumferences of the $i$th handle.

Let $\{\omega_1,\omega_2,\dots, \omega_g\}$ be a normalized basis of $\Omega^1$, the linear space of all holomorphic 1-forms over $\mathbb{C}$. The choice of basis is dependent on the homology basis chosen above; the normalization signifies that
\[
    \int_{a_i} \omega_j = \delta_{ij},\quad i,j = 1,2,\dots,g.
\]
For each curve $\gamma$ in the homology group, we can associate a vector $\lambda_\gamma$ in $\mathbb{C}^g$ by integrating each of the $g$ 1-forms over $\gamma$,
\[
    \lambda_\gamma = \left(\int_\gamma \omega_1,\int_\gamma \omega_2,\dots, \int_\gamma \omega_g \right)
\]
We define a $2g$-real-dimensional lattice $\Lambda$ in $\mathbb{C}^g$,
\[
    \Gamma = \left\{  \sum_{i=1}^g s_i~\lambda_{a_i} + \sum_{j=1}^g t_j~ \lambda_{b_j}, \quad s_i,t_j\in \mathbb{Z} \right\}
\]
\begin{definition}[Jacobian]
The Jacobian of the Riemann surface $S$, denoted $J(S)$, is the compact quotient $\mathbb{C}^g/\Lambda$.
\end{definition}

\begin{definition}[Abel-Jacobi Map] Fix a base point $p_0\in S$. The Abel-Jacobi map is a map $\mu:S\to J(S)$. For every point $p\in S$, choose a curve $c$ from $p_0$ to $p$; the Abel-Jacobi map $\mu$ is defined as follows:
\[
    \mu(p) = \left(\int_{p_0}^q \omega_1, \int_{p_0}^q \omega_2, \dots, \int_{p_0}^q \omega_g \right) ~~\mod \Lambda,
\]
where the integrals are all along $c$.
\end{definition}
It can be shown $\mu(p)$ is well-defined,  that the choice of curve $c$ doesn't not affect the value of $\mu(p)$.

\begin{theorem}[Abel-Jacobian] Let $D$ be an divisor of degree $0$ on $S$, then $D$ is the divisor of a meromorphic function $f$ if and only if $\mu(D)=0$ in the Jacobian $J(S)$.
\end{theorem}

\subsection{Quad-Meshes and Meromorphic Quartic Forms}
\label{subsec:quad_differential}

We summarize the intrinsic relation between a quad-mesh and a meromorphic quartic differential.

\begin{definition}[Quadrilateral Mesh] Suppose $\Sigma$ is a topological surface, $\mathcal{Q}$ is a cell partition of $\Sigma$, if all cells of $\mathcal{Q}$ are topological quadrilaterals, then we say $(\Sigma,\mathcal{Q})$ is a quadrilateral mesh.
\end{definition}
On a quad-mesh, the \emph{topological valence} of a vertex is the number of faces adjacent to the vertex.
\begin{definition}[Singularity] Suppose $(S,\mathcal{Q})$ is a quadrilateral mesh. If the topological valence of an interior vertex is $4$, then we call it a \emph{regular vertex}, otherwise a \emph{singularity}; if the topological valence of a boundary vertex is $2$, then we call it a \emph{regular boundary vertex}, otherwise a \emph{boundary singularity}. The index of a singularity is defined as follows:
\[
    \text{Ind}(v_i) = \left\{
    \begin{array}{lcl}
    4-\text{val}(v_i) & v_i\not\in \partial (S,\mathcal{Q})\\
    2-\text{val}(v_i) & v_i\in \partial (S,\mathcal{Q})\\
    \end{array}
    \right.
\]
where $\text{Ind}(v_i)$ and $\text{val}(v_i)$ are the index and the topological valence of $v_i$.
\end{definition}

\begin{theorem}[Qaud-Mesh to Meromrophic Quartic Differential]
Suppose $(\Sigma,\mathcal{Q})$ is a closed quadrilateral mesh, then
\begin{enumerate}
\item the quad-mesh $\mathcal{Q}$ induces a conformal atlas $\mathcal{A}$, such that $(\Sigma,\mathcal{A})$ form a Riemann surface, denoted as $S_Q$.
\item the quad-mesh $\mathcal{Q}$ induces a quartic differential $\omega_Q$ on $S_Q$. The valence-$k$ singular vertices correspond to poles or zeros of order $k-4$. Furthermore, the trajectories of $\omega_Q$ are finite.
\end{enumerate}
\label{thm:quad_differential}
\end{theorem}

\begin{theorem}[Quartic Differential to Quad-Mesh]
Suppose $(\Sigma,\mathcal{A})$ is a Riemann surface, $\omega$ is a meromorphic quartic differential with finite trajectories, then $\omega$ induces a quadrilateral mesh $\mathcal{Q}$, such that the poles or zeros with order $k$ of $\omega$ corresponds to the singular vertices of $\mathcal{Q}$ with valence $k+4$.
\label{thm:differential_quad}
\end{theorem}

\begin{theorem}[Quad-mesh singularity Abel-Jacobian condition]Suppose $\mathcal{Q}$ is a closed quadrilateral mesh, $S_Q$ is the induced Riemann surface, $\omega_Q$ is the induced meromorphic quadric form. Assume $\omega_0$ is an arbitrary holomorphic 1-form on $S_Q$, then
\begin{equation}
        \mu((\omega_Q) - 4(\omega_0)) = 0\quad \mod \Lambda
        \label{eqn:abel_condition}
\end{equation}
in the Jacobian $J(S_Q)$.
\label{thm:Abel_Jacobian_condition}
\end{theorem}


\section{Computational Algorithms}
\label{sec:algorithm}
This section explains the algorithm in details. The input surface is represented as a triangle mesh $\Sigma$; the output is a meromorphic quartic differential $\omega$, and the flat metric with cone singularities at the poles and zeros induced by $\omega$. The pipeline of the algorithm is as follows:
\begin{enumerate}[1.]
\item Compute the \emph{homology} group generators of $\Sigma$, $\{a_1,\cdots,a_g; \  b_1,\cdots,$ $b_g\}$;
\item Compute the dual \emph{holomorphic} 1-form basis $\{ \varphi_1, \cdots, \varphi_g \}$; Construct a \emph{holomorphic} differential $\varphi$ on the Riemann surface $\Sigma$ through a linear combination of basis $\{\varphi_k\}_{k=1}^g$, locate the zeros of $\varphi$;
\item Compute the period matrix $(A,B)$ of surface and construct the lattice $\Gamma$, Jacobian variet $J(\Sigma)$;
\item Compute Abel-Jacobi map of a given divisor $D$ in the  Jacobian variet $J(\Sigma)$;
\item Optimize the divisor $D$ to satisfy the Abel-Jacobian condition;
\item Compute the flat metric with cone singularities at the divisor $D$ by surface \emph{Ricci Flow};
\item Compute the cut-graph connecting all singularities; slice the surface along the cut-graph; And isometrically immerse the surface into complex plane; the immersion pulls $(dz)^4$ back to the surface and produces a meromorphic quartic differential $\omega$.
\item Trace the critical horizontal and vertical trajectories of $\omega$, namely isoparametric curves through singularities, to generate a T-mesh and partition the surface into rectangular patches.
\end{enumerate}
In the following, we explain every step in details. Each subsection corresponds to one step.

\subsection{Homology group basis}
\label{sec:homology_group_basis}

In practice, we compute a special set of canonical homology group basis, the tunnel loops $\{a_i\}$ and handle loops $\{b_i\}$, such that each $a_i$ and $b_i$ intersect each other at one point.
Our algorithm is mainly based on the work of Dey et al\cite{}, which avoid tetrahedral tessellation and modification of the original triangle mesh. The algorithm utilizes the concept of \emph{reeb graph} and the linking number to produce different sets of \emph{homology} basis.

\begin{figure}[h!]
\centering
\begin{tabular}{cc}
\includegraphics[width=0.45\textwidth]{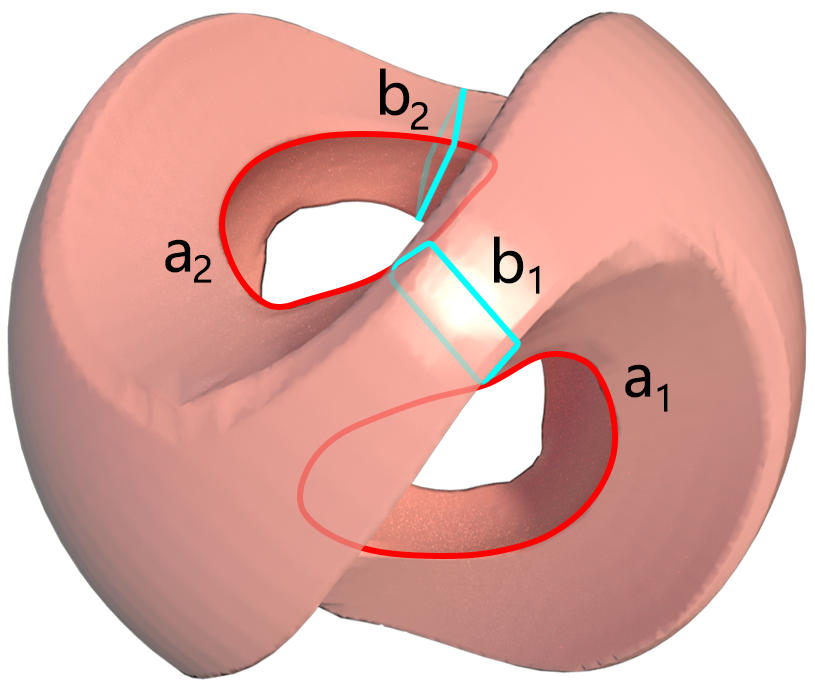}&
\includegraphics[width=0.45\textwidth]{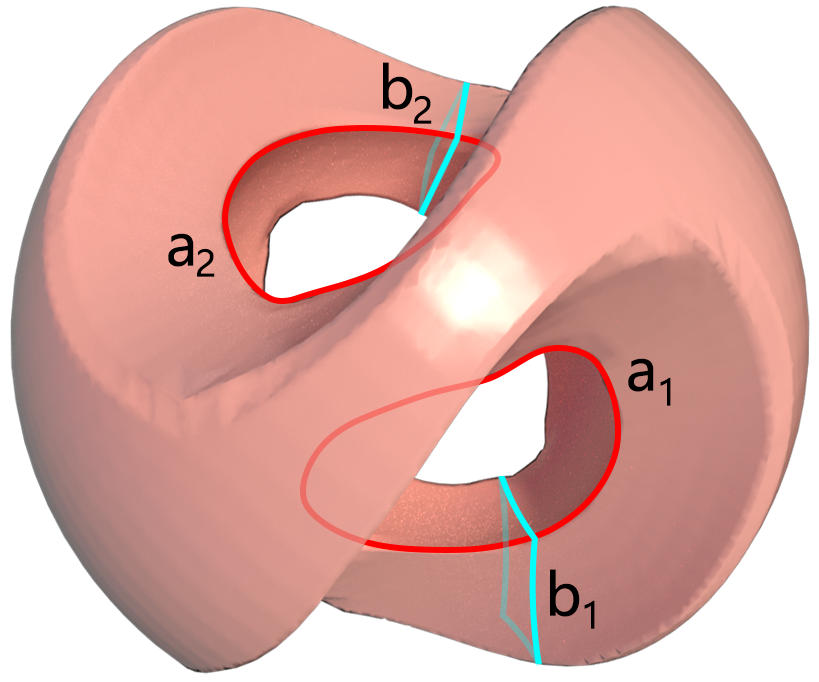}
\end{tabular}
\caption{\textbf{Sculpt} model. \textbf{Left}: the handle loop $b_1$ intersects both tunnel loops $a_1$ and $a_2$, where we call the handle loop $b_1$ illegal. \textbf{Right}: each handle loop $b_i$ only intersects its conjugate $a_i$ once.}
\label{fig:sculpt_loops}
\end{figure}

As shown in Fig.~\ref{fig:sculpt_loops} left frame, the algorithm may generate homology basis which doesn't satisfy the intersection condition,
\begin{equation}
    a_i \cdot b_j = \delta_{ij},~a_i\cdot a_j=0,~b_i\cdot b_j=0, \quad i,j=1,\dots, g,
    \label{eqn:intersection_condition}
\end{equation}
where $\alpha\cdot \beta$ represents the algebraic intersection number between $\alpha$ and $\beta$, $g$ is the genus of the mesh.
We compute the algebraic intersection between the tunnel loops and handle loops, if the intersection condition \ref{eqn:intersection_condition} is violated, we randomly reset the height function used for constructing  reeb graph, and obtain a new set of handle loops and tunnel loops. After several iterations, we can get a set of canonical homology group basis, as shown in Fig.~\ref{fig:sculpt_loops} right frame.

\subsection{Holomorphic 1-form Basis}
\label{sec:holomorphic1form_basis}

\begin{figure}[h!]
\centering
\begin{tabular}{rl}
\includegraphics[width=0.4\textwidth]{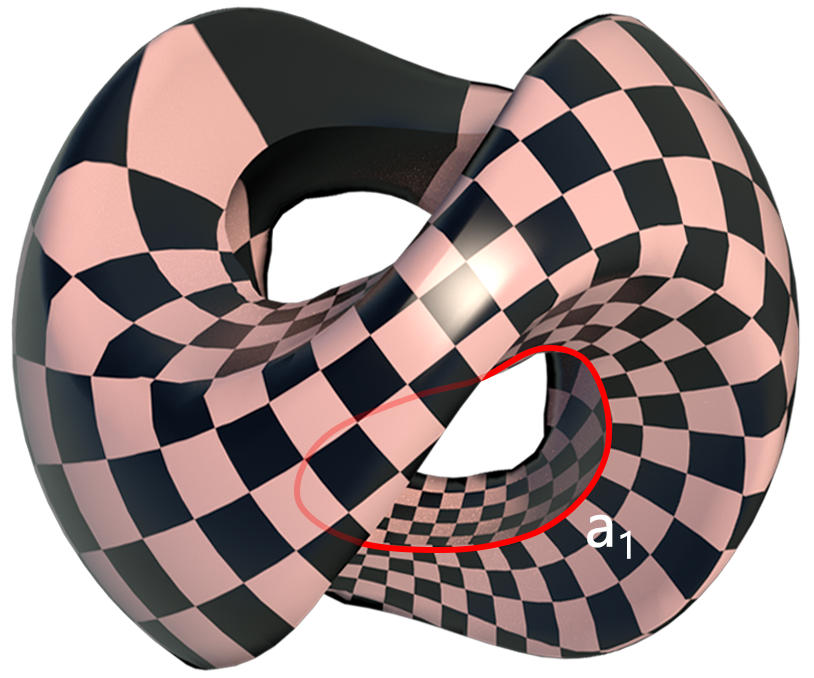}&
\includegraphics[width=0.4\textwidth]{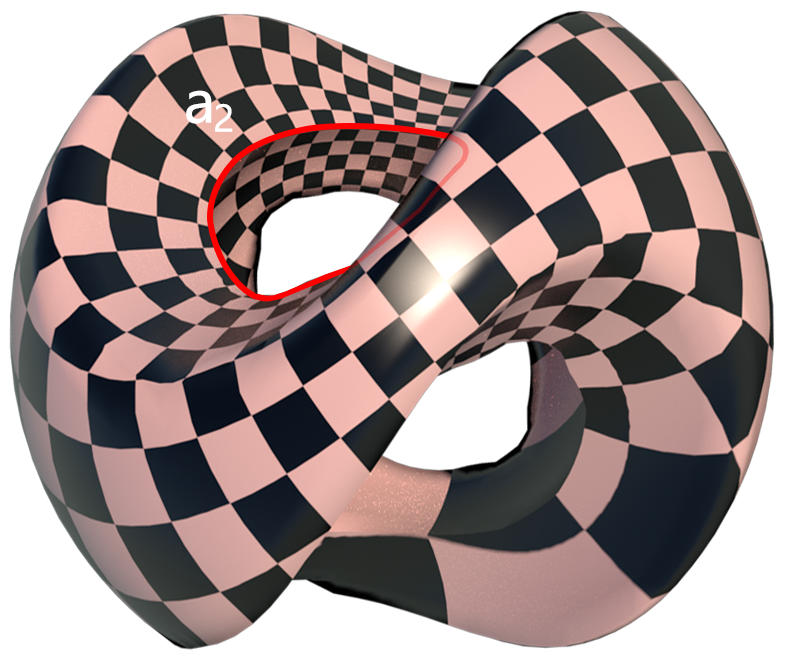}\\
\includegraphics[width=0.4\textwidth]{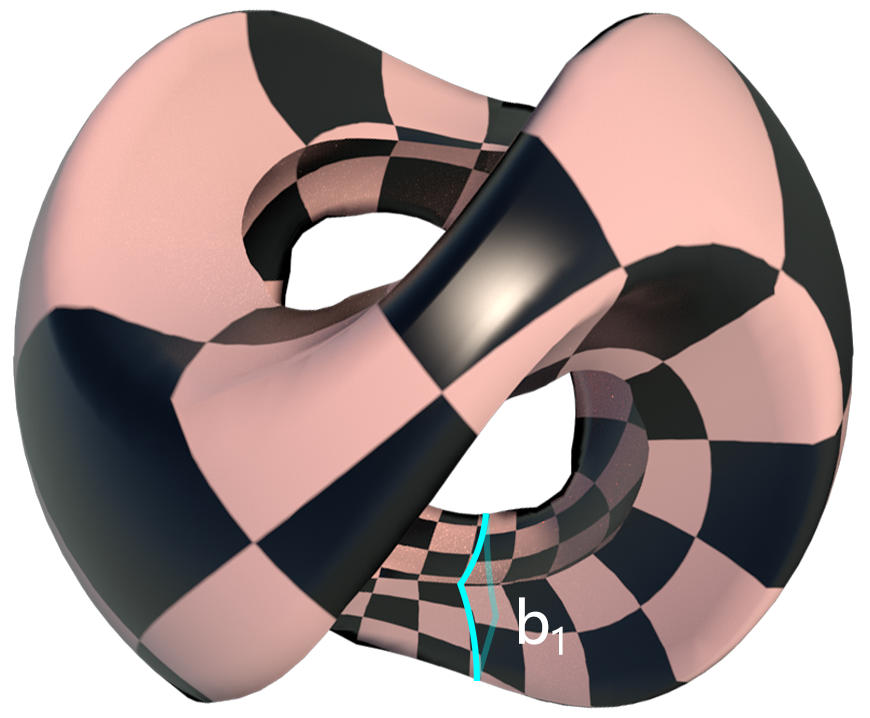}&
\includegraphics[width=0.4\textwidth]{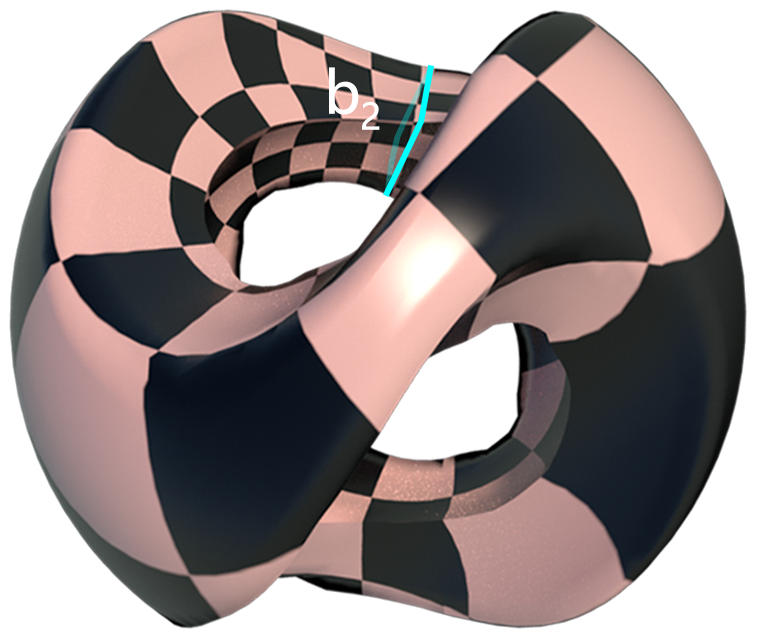}\\
\end{tabular}
\caption{A \emph{holomorphic} 1-form basis on a genus two surface, the \textbf{Sculpt} model.}
\label{fig:sculpt_holo1form_basis}
\end{figure}

The algorithm of computing \emph{holomorphic} 1-form is based of the work of Gu et al \cite{}, which is based on Hodge theory. \textbf{Step 1}, for each loop $\gamma$, we slice the mesh $M$ along $\gamma$ to get an open mesh $\bar{M_\gamma}$ with boundaries $\partial M_\gamma=\gamma^+-\gamma^-$, then we construct a function
\[
    g_\gamma(v_i) = \left\{
\begin{array}{ll}
    1& v_i \in \gamma^{+}\\
    0& v_i \in \gamma^{-}\\
    \text{rand}&\text{otherwise}
\end{array}
    \right.
\]
Then the discrete 1-form $\lambda_{\gamma}=dg_\gamma$ is a closed 1-form.
In this way, we construct a set of cohomology group basis $\lambda_{a_1},\lambda_{b_1},\cdots, \lambda_{a_g},\lambda_{b_g}$. \textbf{Step 2}, for each closed 1-form $\lambda$, we construct a function $f:M\to\mathbb{R}$, such that $\lambda+df$ is harmonic, namely the function $f$ satisfies the Poisson equation $\Delta f = -\delta \lambda$. In this way, we diffuse cohomology basis to harmonic 1-form group basis, denoted as $\omega_{a_1},\omega_{b_1},\cdots,\omega_{a_g},\omega_{b_g}$. \textbf{Step 3}, each harmonic 1-form $\omega$ is equivalent to a curl-free vector field on $M$, we rotate the vector field by $\frac{\pi}{2}$ about the normal to the surface to obatin a divergence free vector field, which is equivalent to another harmonic 1-form ${}^*\omega$. The pair $\omega+\sqrt{-1}{}^*\omega$ is a holomorhic 1-form. In this way, we construct the holomorphic 1-form basis $\{\varphi_{a_1},\varphi_{b_1},\dots,\varphi_{a_g},\varphi_{b_g}\}$, where
\[
    \varphi_{\gamma}= \omega_\gamma + \sqrt{-1}{}^*\omega_\gamma, \gamma\in \{a_1,\dots,a_g,b_1,\dots,b_g\}.
\]
According to the Riemann-Roch theory, the above set of \emph{holomorphic} 1-forms span the linear space of all holomorphic 1-forms $\Omega$, namely
for any $\varphi\in \Omega$,
\[
\varphi = \sum_{k=1}^g \alpha_k \varphi_{a_k} + \sum_{l=1}^g \beta_l \varphi_{b_l},
\]
where the $\alpha_k$, $\beta_l$ are real linear combination coefficients.

In practice, in order to compute the zeros of $\varphi$ more accurately, we choose the linear combination coefficients, such that the conformal factor function of $\varphi$ is as uniform as possible. In our implementation, we assign all $\alpha_k$'s and $\beta_l$'s to be $1$. Heuristically, the resulting holomorphic 1-form meets our accuracy requirement.

\subsection{Period matrix and Lattice}
\label{sec:period_matrix}
We can further construct a set of holomorphic 1-form basis $\{\varphi_1,\varphi_2,\dots,\varphi_g\}$, such that
\[
\int_{a_i} \varphi_j = \delta_{ij}, i,j = 1,2,\dots,g.
\]
Then for each $\gamma$ in the homology basis, we construct a $g$ dimensional vector $\lambda_\gamma \in \mathbb{C}^g$,
\[
    \lambda_\gamma = \left(\int_\gamma \varphi_1, \int_\gamma \varphi_2,\cdots, \int_\gamma \varphi_g\right)
\]
The period matrix $(A,B)$ can be constructed as
\begin{equation}
A =
\begin{bmatrix}
\int_{a_1} \varphi_{1} & \int_{a_2} \varphi_{1} & \cdots & \int_{a_g} \varphi_{1}\\
\int_{a_1} \varphi_{2} & \int_{a_2} \varphi_{2} & \cdots & \int_{a_g} \varphi_{2}\\
\vdots  &\vdots  &\ddots &\vdots \\
\int_{a_1} \varphi_{g} & \int_{a_2} \varphi_{g} & \cdots & \int_{a_g} \varphi_{g}\\
\end{bmatrix}
~B =
\begin{bmatrix}
\int_{b_1} \varphi_{1} & \int_{b_2} \varphi_{1} & \cdots & \int_{b_g} \varphi_{1}\\
\int_{b_1} \varphi_{2} & \int_{b_2} \varphi_{2} & \cdots & \int_{b_g} \varphi_{2}\\
\vdots  &\vdots  &\ddots &\vdots \\
\int_{b_1} \varphi_{g} & \int_{b_2} \varphi_{g} & \cdots & \int_{b_g} \varphi_{g}\\
\end{bmatrix}
\end{equation}
by our construction $A$ is the $g\times g$ identity matrix.
Then we construct a lattice $\Gamma$ in $\mathbb{C}^g$,
\[
\Gamma = \left\{
\sum_{k=1}^g (s_k \lambda_{a_k} + t_k \lambda_{b_k},\quad s_k, t_k\in \mathbb{Z}
\right\}
\]

\subsection{Abel-Jacobi Map}
\label{sec:Abel_Jacobi}

Given a canonical homology group basis£¬ we slice the surface along the basis to obtain a topological disk $\bar{M}$. Fix a base point in the interior of $\bar{M}$,$p_0\in \bar{M}$, for any point $p\in M$, we can choose arbitrarily a path $\gamma\subset \bar{M}$ connecting $p$ and $p_0$, the Abel-Jacobi map $\mu:M\to J(M)$, $J(M)=\mathbb{C}^g/\Gamma$, is defined as
\[
    \mu(p) = \Phi(p)~\mod~\Gamma,
\]
where
\begin{equation}
\Phi(p) = \left(\int_\gamma \varphi_1, \int_\gamma \varphi_2,\cdots, \int_\gamma \varphi_g\right)^T.
\label{eqn:Phi}
\end{equation}
Similarly, given a divisor $D=\sum_{i=1}^n n_i p_i$,
\[
    \mu(D) = \sum_{i=1}^n n_i \mu(p_i) = \sum_{i=1}^n n_i\left(\int_{p_0}^{p_i} \varphi_1, \int_{p_0}^{p_i} \varphi_2,\cdots, \int_{p_0}^{p_i} \varphi_g\right)^T ~\mod~\Gamma.
\]
Abel-Jacobi condition claims that if $D$ is a principle divisor, then $\mu(D)$ is $0$, namely
\begin{equation}
\Phi(D) - A
\begin{bmatrix}
s_1\\
\vdots\\
s_g
\end{bmatrix}
- B
\begin{bmatrix}
t_1\\
\vdots\\
t_g
\end{bmatrix}
=
\begin{bmatrix}
0\\
\vdots\\
0
\end{bmatrix}
\label{eqn:Abel_Jacobi_Condition}
\end{equation}
By expansion, we obtain the equation
\begin{equation}
\begin{split}
\centering
\left[
\begin{array}{c}
\mathop{\sum}_{i=1}^{n} n_i \bigintss_{\gamma_i} \varphi_1 \\
\mathop{\sum}_{i=1}^{n} n_i \bigintss_{\gamma_i} \varphi_2 \\
\quad \vdots \\
\mathop{\sum}_{i=1}^{n} n_i \bigintss_{\gamma_i} \varphi_g
\end{array}
\right]
\
-
\
\begin{bmatrix}
\mathop{\sum}_{k = 1}^{g} (s_k \bigintss_{a_k} \varphi_1 + t_k \bigintss_{b_k} \varphi_1) \\
\mathop{\sum}_{k = 1}^{g} (s_k \bigintss_{a_k} \varphi_2 + t_k \bigintss_{b_k} \varphi_2) \\
\vdots \\
\mathop{\sum}_{k = 1}^{g} (s_k \bigintss_{a_k} \varphi_g + t_k \bigintss_{b_k} \varphi_g)
\end{bmatrix}
=
\begin{bmatrix}
0 \\
0\\
\vdots \\
0
\end{bmatrix},
\end{split}
\label{eqn:abel_jacobian_map_expand_2}
\end{equation}
where $s_k$, $t_k$ are integers, $\gamma_i$ is the path connecting $p_0$ and $p_i$ in $\bar{M}$.

\subsection{\textbf{Abel-Jacobian Condition Optimization System}}
\label{sec:optimization_system}

Suppose we are given an initial divisor $D_0$, if it doesn't satisfy the Gauss-Bonnet condition, namely $\deg(D_0)\neq 8g-8$, then we can add extra poles or zeros to modify $D_0$ to $D=\sum_{i=1}^k n_i p_i$, such that $\deg(D)=8g-8$. We choose a holomorphic 1-form $\varphi$.

First, we determine the integer coefficients $s_k$ and $t_k$ in Abel-Jacobi condition (\ref{eqn:abel_jacobian_map_expand_2}) by minimizing the norm
\begin{equation}
\min_{s_k,t_k\in\mathbb{Z}}\left\|\Phi(D) - \sum_{k=1}^g s_k \lambda_{a_k} - \sum_{k=1}^g t_k \lambda_{b_k} - \Phi(4(\varphi))\right\|^2,
\label{eqn:integer_program}
\end{equation}
this can be accomplished by standard integer programming \cite{}.

Second, once the integer coefficients $s_k, t_k$, $k=1,2,\dots,g$ are set, we further minimize the
squared norm of $\mu(D)$ with respect to the positions of poles and zeros,
\[
    \min_{p_1,\dots p_k\in M} \left\|\Phi(D) - \sum_{k=1}^g s_k \lambda_{a_k} - \sum_{k=1}^g t_k\lambda_{b_k} - \Phi(4(\varphi))\right\|^2.
\]
Let $d\in \mathbb{C}^g$ be
\[
    d = \sum_{k=1}^g s_k \lambda_{a_k} - \sum_{k=1}^g t_k\lambda_{b_k} - \Phi(4(\varphi)),
\]
then the above energy becomes
\[
E(p_1,\dots,p_k) := \sum_{j=1}^g  \left\|\sum_{i=1}^k n_i \int_{p_0}^{p_i} \varphi_j - d_j  \right\|^2.
\]
For each point $p_i$, we choose a local neightborhood $\Delta_i$ of $p_i$, with local parameter $z_i$, then the holomorphic 1-form $\omega_j$ has local representation,
\[
    \varphi_j = h_{j}^i(z_i) dz_i,
\]
where $h_j^i(z_i)$ is a holomorphic function defined on $\Delta_i$.
\begin{equation}
\frac{\partial E}{\partial p_i}=\sum_{j=1}^g \left[ n_i h_j^i(p_i) \left(\sum_{i=1}^k n_i \int_{p_0}^{p_i} \bar{\varphi}_j -\bar{d}_j\right) +n_i \bar{h}_j^i(p_i) \left(\sum_{i=1}^k n_i \int_{p_0}^{p_i} \varphi_j-d_j\right)
\right],
\label{eqn:gradient}
\end{equation}
we can use gradient descent method to minimize $\|\mu(D)\|^2$. In practice, we choose the triangle face containing $p_i$ as $\Delta_i$. We isometrically embed $\Delta_i$ onto the plane, and the planar coordinates give local parameter $z_i$. $\omega_j$ can be represented as a complex linear function on $\Delta_i$, which is $h_j^i(z_i)$. In this way, we can minimize the squared norm of $\mu(D)$.

The algorithm is presented briefly in Alg.~\ref{alg:solve_optimization_system}.
\begin{algorithm}[h]
\caption{Optimize a Divisor to Satisfy the Abel-Jacobi Condition}
\renewcommand{\algorithmicrequire}{\textbf{Input:}}
\renewcommand{\algorithmicensure}{\textbf{Output:}}
\label{alg:solve_optimization_system}
\begin{algorithmic}[1]
\REQUIRE Closed mesh $M$; A group of singularities $D$; A \emph{holomorphic} 1-form; Precision threshold $\varepsilon$.
\ENSURE Optimized divisor $D$ Abel-Jacobian condition.
\IF{$D$ doesn't satisfy Gauss-Bonnet Condition}
\STATE Locate the vertices on $M$ with local maximal Gaussian curvature as poles, or with local minimal curvature as zeros;
\STATE Add these vertices to the divisor $D$, such that $D$ satisfies the Gauss-Bonnet condition.
\ENDIF
\STATE Locate the zeros of $\varphi$ to obtain the divisor $(\varphi)$;
\STATE Compute $\Phi(D)$ and $\Phi(4(\varphi))$ using Eqn.~\ref{eqn:Phi};
\STATE Compute the Abel-Jacobi $\mu(D-4(\varphi))$map by optimization using integer programming Eqn.\ref{eqn:integer_program};
\WHILE{\;$\|\mu(D-4(\varphi))\|^2 > \varepsilon$\;}
\FOR {All each pole and zero $p_i$ in $D$ }
\STATE Locate the face $\Delta_i$ containing $p_i$;
\STATE Compute the local representation $\varphi_j(z_i) = h_j^i(z_i) dz_i$;
\STATE Compute the gradient of the energy Eqn.~\ref{eqn:gradient};
\ENDFOR
\STATE Update the poisitions of the singularities $p_i \leftarrow p_i - \partial \nabla E/\partial p_i$;
\STATE Recompute the Abel-Jacobi map $\mu(D-4(\varphi))$;
\ENDWHILE
\RETURN The divisor $D$.
\end{algorithmic}
\end{algorithm}

\begin{figure}[h!]
    \centering
\begin{tabular}{cc}
\includegraphics[height=0.55\textwidth]{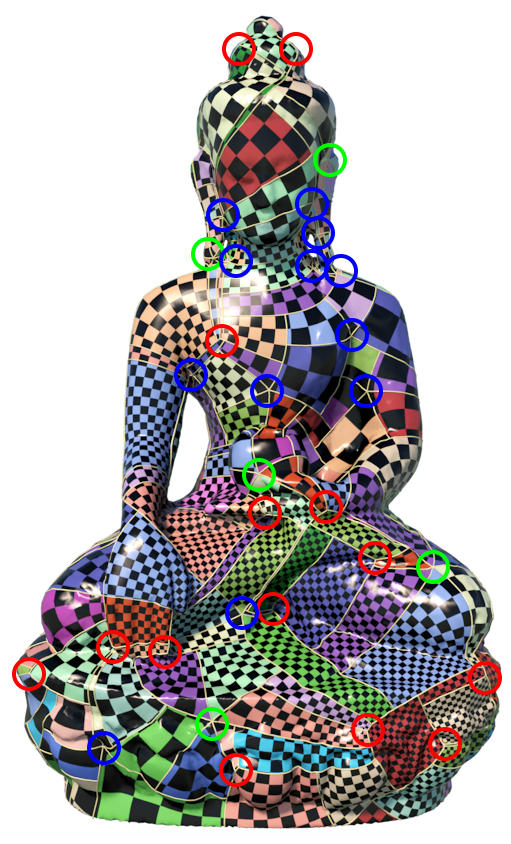}&
    \includegraphics[height=0.55\textwidth]{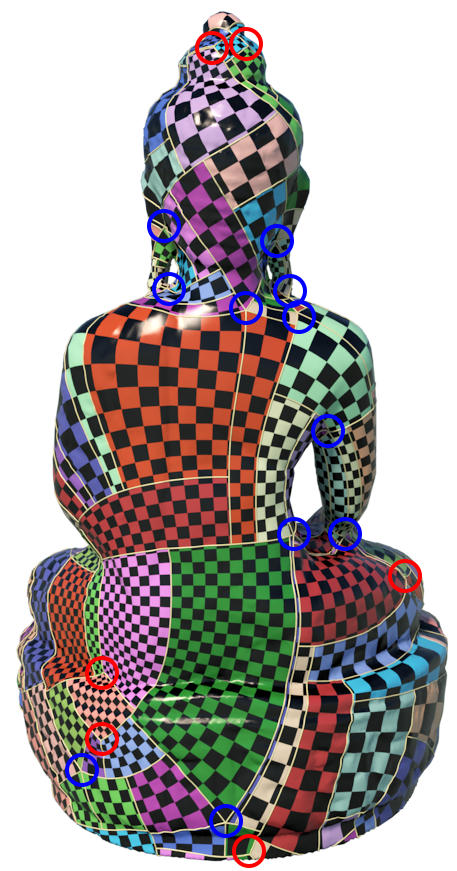}
\end{tabular}
    \caption{The singularities of the Buddha surface.}
    \label{fig:buddha_singularities}
\end{figure}
Fig.~\ref{fig:buddha_singularities} shows the singularities on the Buddha surface satisfying the Abel-Jacobi condition.

\subsection{Discrete Surface Ricci Flow}
Once the divisor $D=\sum_{i=1}^k n_i p_i$ is obtained, we can set the target curvature as
\[
    \bar{K}(v_i) = \left\{
    \begin{array}{ll}
    (4-n_i)\frac{\pi}{2}     &  v_i \in D\\
        0 & \text{otherwise}
    \end{array}
    \right.
\]
For each vertex $v_i \in M$, we set the initial conformal factor as $u_i=0$. Then the edge length is given by vertex scaling, for edge $e_{ij}=[v_i,v_j]$, its length is given by
\[
    l_{ij} = e^{u_i}\beta_{ij} e^{u_j},
\]
where $\beta_{ij}$ is the initial edge length. The corner angles are calculated using Euclidean cosine law,
\[
    \theta_{k}^{ij} = \cos^{-1} \frac{l_{ik}^2 + l_{jk}^2 - l_{ij}^2}{2l_{ik}l_{jk}},
\]
the discrete Gaussian curvature is given by
\[
    K(v_i) = \left\{
    \begin{array}{rl}
    2\pi - \sum_{jk} \theta_i^{jk}     & v_i \not\in \partial M \\
    \pi -  \sum_{jk} \theta_i^{jk}  & v_i \in \partial M
    \end{array}
    \right.
\]
The discrete Ricci energy is defined as
\[
    E(u_1,\dots,u_n)=\int^{(u_1,\dots,u_n)} \sum_{i=1}^n (\bar{K}_i - K_i ) du_i.
\]
The gradient of the energy is given by
\[
    \nabla E=(\bar{K}_1-K_1,\bar{K}_2-K_2,\cdots, \bar{K}_n-K_n)^T.
\]
The Hessian matrix is given by the cotange edge weight
\[
\frac{\partial^2 E}{\partial u_i \partial u_j} =
\left\{
\begin{array}{lr}
(\cot \theta_{k}^{ij} + \cot \theta_{l}^{jl})/2& e_{ij}\not\in \partial M\\
\cot \theta_{k}^{ij}/2 & e_{ij}\in \partial M
\end{array}
\right.
\]
and
\[
\frac{\partial^2 E}{\partial u_i^2} = -\sum_{j\neq i} \frac{\partial^2 E}{\partial u_i \partial u_j}.
\]
We can use Newton's method to optimize the Ricci energy, during the optimization, we update the triangulation to be Delaunay all the time. The convergence is proven in the work \cite{}. We can compute holonomy using the resulting Riemannian metric.

\subsection{Isometric Immersion and Meromorphic Quartic Differential}
We have obtain a set of canonical homology group basis $\{a_1,\dots,a_g,b_1,\dots,b_g\}$. The union of the basis form a cut graph $\Gamma$ of the mesh. For each pole or zero $p_i$ in $D$, we find a shortest path $\gamma_i$ connecting $p_i$ to the cut graph, furthermore, all such shortest paths $\gamma_i$'s are disjoint. Then we slice $M$ along the cut graph and the shortest paths, $\Lambda \bigcup \{\bigcup_{i=1}^k \gamma_i\}$, to obtain a topological disk $\tilde{M}$.

Then we flatten $\tilde{M}$ face by face using the metric obtained by the discrete surface Ricci flow. This produces an immersion of $\varphi:\tilde{M}\to\mathbb{C}$. On the complex plane, there is a canonical differential $dz^4$, the pull back $\varphi^* dz^4$ is a meromorphic quartic differential defined on $M$. We can use $\tau$ as a parameterization, and use checker board texture mapping to visualize the quartic differential.

\begin{figure}[t!]
    \centering
\begin{tabular}{cccc}
    \includegraphics[height=0.35\textwidth]{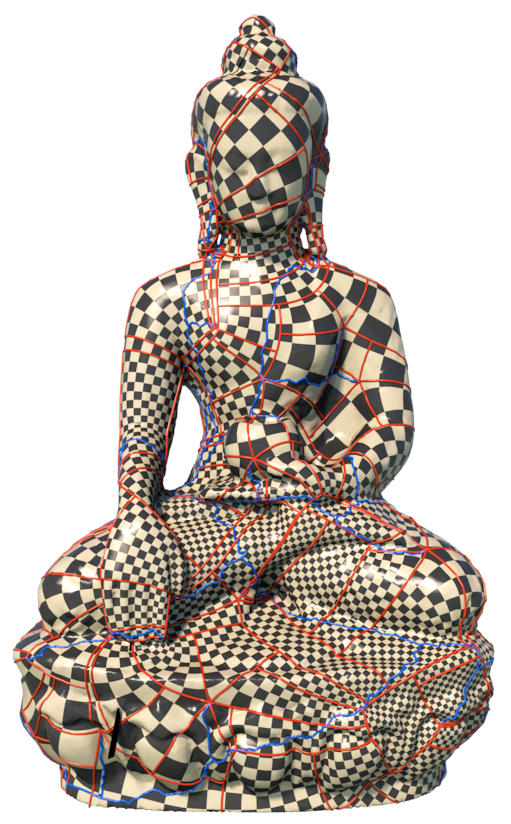}&
    \includegraphics[height=0.35\textwidth]{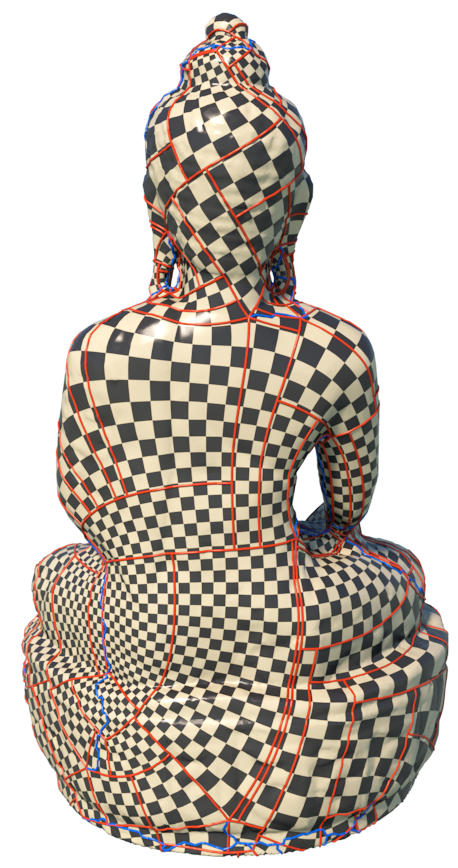}&
        \includegraphics[height=0.35\textwidth]{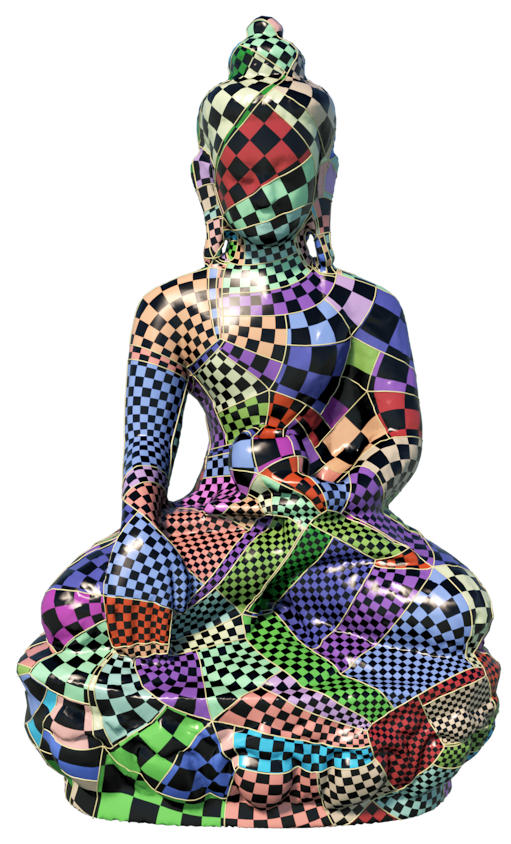}&
    \includegraphics[height=0.35\textwidth]{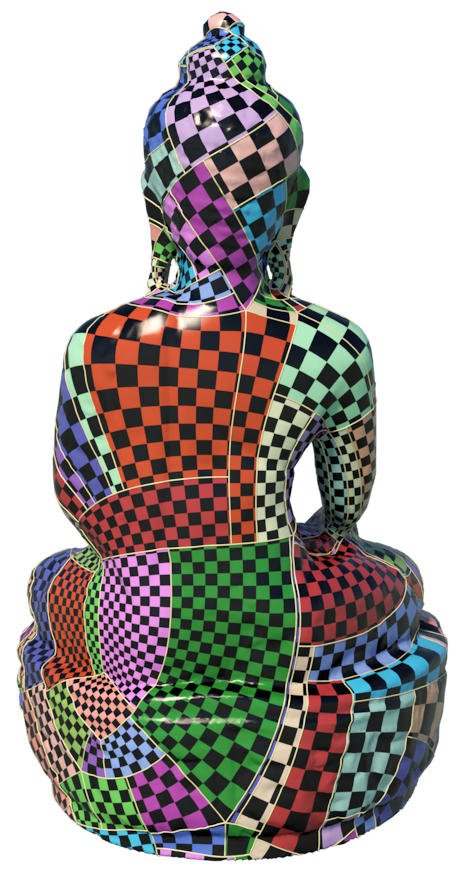}\\
\end{tabular}
    \caption{In the left two frames, the red curves form the motor-graph, the blue curves are the original cut graph. The right two frames show the T-Mesh of the Buddha surface.}
    \label{fig:motorgraph_TMesh}
\end{figure}

\subsection{T-Mesh Generation}
We trace the critical trajectories of the meomorphic quadratic differential $\varphi^*dz^4$, denoted as $\{\gamma_1(s_1),\gamma_2(s_2),\cdots,\gamma_n(s_n)\}$, where $s_k$ is the arc length parameter of $\gamma_k$, their images $\varphi(\gamma_k)$'s are the horizontal and vertical lines through the zeros and poles on the parameter plane. If $\gamma_i(s_i)$ intersects $\gamma_j(s_j)$ at $p$, if $s_i<s_j$, then $\gamma_j$ stops at $p$, $\gamma_i$ continues. This procedure will generate the \emph{motor graph} on the surface, as shown in the left two frames in  Fig.~\ref{fig:motorgraph_TMesh}.

\begin{figure}[t!]
    \centering
\begin{tabular}{cc}
    \includegraphics[height=0.3\textwidth]{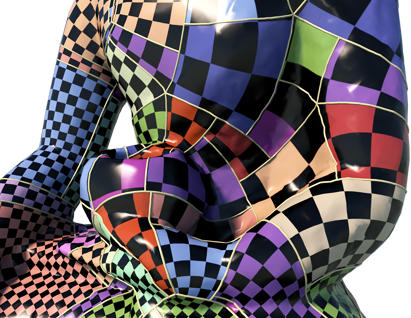}&
    \includegraphics[height=0.35\textwidth]{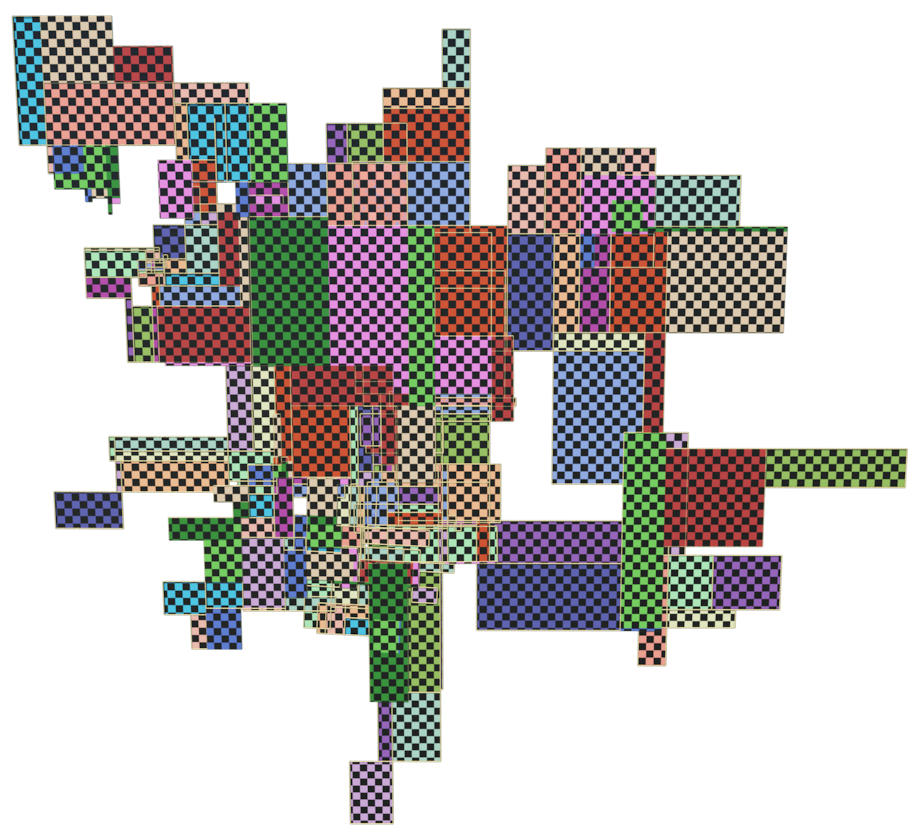}\\
    (a) zoomed in & (b) planar images\\
\end{tabular}
    \caption{Each surface patch is parameterized to a planar rectangle.}
    \label{fig:buddha_parameterization}
\end{figure}

The surface is partitioned into rectangular patches as shown in the right two frames of Fig.~\ref{fig:motorgraph_TMesh}. Each surface patch is parameterized to a planar rectangle, as shown in Fig.~\ref{fig:buddha_parameterization}. The corresponding surface patch and the planar rectangle are rendered using the same color.

\section{Experimental Results}
\label{sec:experiments}

In this section, we briefly report our experimental results. All the experiments were conducted on a PC with 1.60GHz Intel(R) core(TM) i5-8250U CPU, 1.60GB RAM and 64-bit Windows 10 operating system. The running time is reported in table \ref{tab:Runing_time}.

\subsection{T-Mesh Generation}
\begin{figure}
    \centering
\begin{tabular}{cc}
    \includegraphics[width=0.5\textwidth]{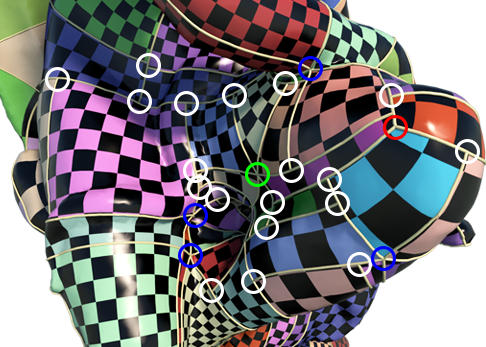}&
    \includegraphics[width=0.5\textwidth]{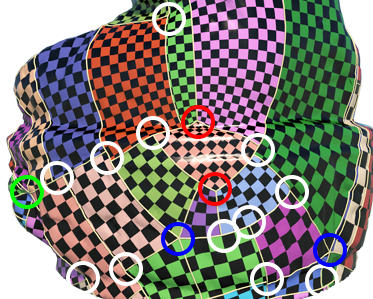}\\
\end{tabular}
    \caption{Singularities.}
    \label{fig:singularities}
\end{figure}

The singularities and the resulting T-meshes are illustrated in the figures. As shown in Fig.~\ref{fig:singularities}, the singularities surrounded by red, blue and green circles represent the indices $+1$, $-1$ and $-2$ respectively. The points surrounded by white circles are the T-junctions of the T-mesh. Different surface patches are color-encoded differently. By carefully examining the texture patterns in Fig.~\ref{fig:singularities}, we can see that the adjacent patches differ by horizontal and vertical translations composed with rotations by angle $k\frac{\pi}{2}$, $k\in \mathbb{Z}$. Therefore, we can construct T-Splines on these T-meshes directly.

\begin{figure}
    \centering
\begin{tabular}{cc}
    \includegraphics[height=0.55\textwidth]{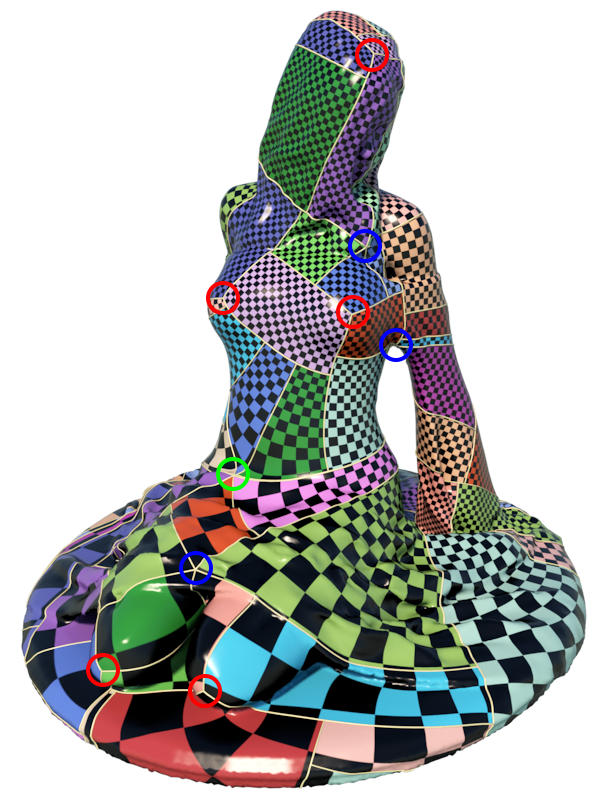}&
    \includegraphics[height=0.55\textwidth]{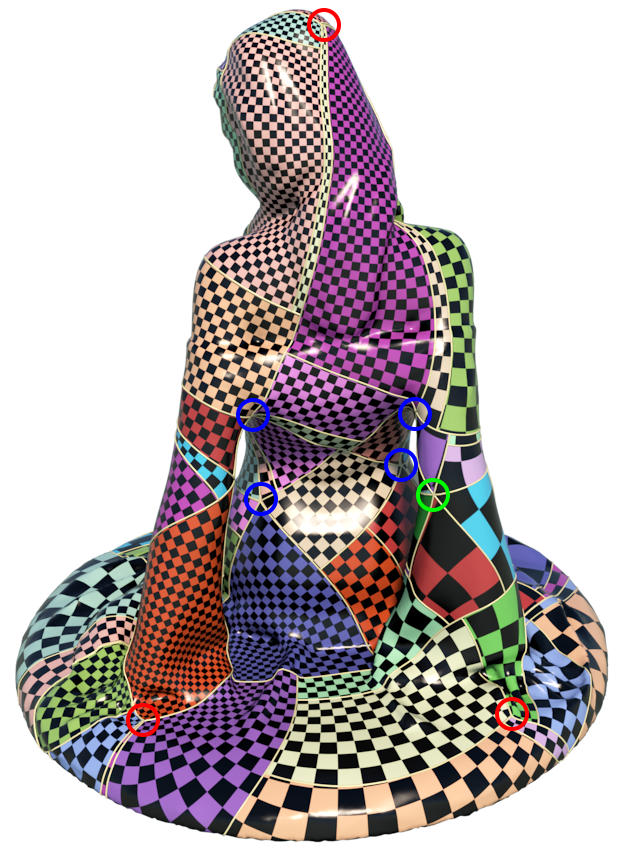}\\
    (a) front view & (b) back view
\end{tabular}
    \caption{Singularities and the T-Mesh of the Loveme model.}
    \label{fig:examples_1}
\end{figure}

\begin{figure}
    \centering
\begin{tabular}{ccc}
         \includegraphics[height=0.60\textwidth]{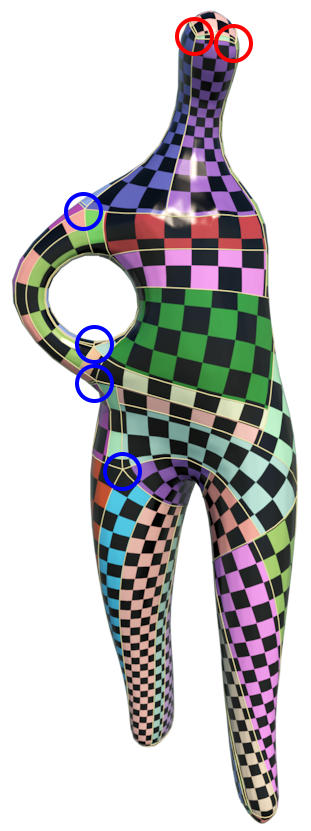}&
    \includegraphics[height=0.60\textwidth]{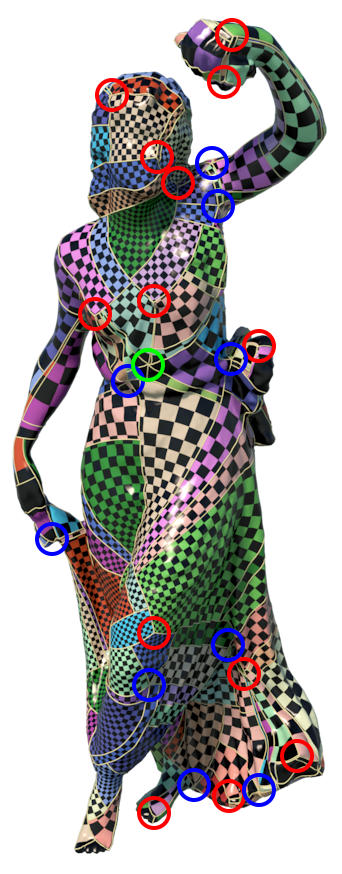}&
    \includegraphics[height=0.60\textwidth]{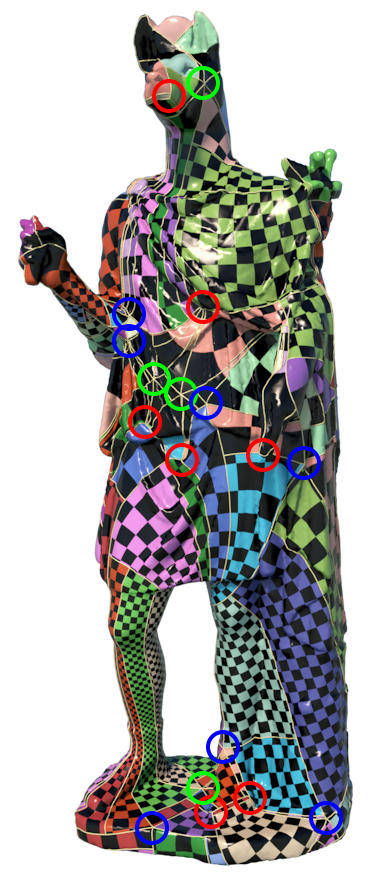}\\
    Ornament& Dancer & Hermanubis\\
\end{tabular}
    \caption{Singularities and the T-meshes of high genus surfaces.}
    \label{fig:examples_2}
\end{figure}

\begin{figure}
    \centering
\begin{tabular}{ccc}
        \includegraphics[height=0.60\textwidth]{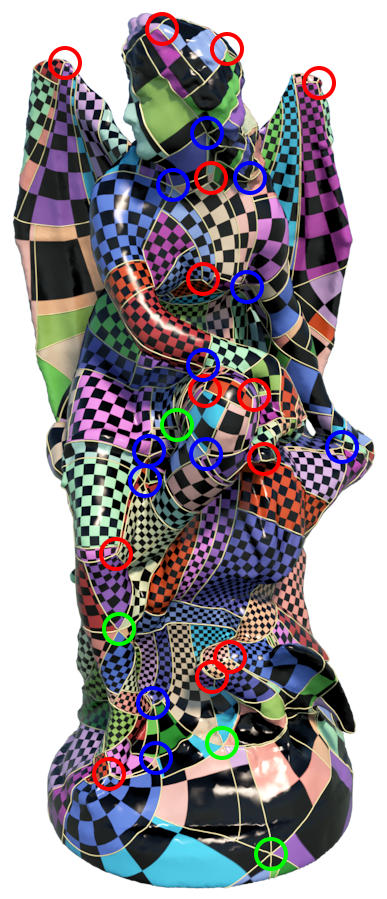}&
        \includegraphics[height=0.60\textwidth]{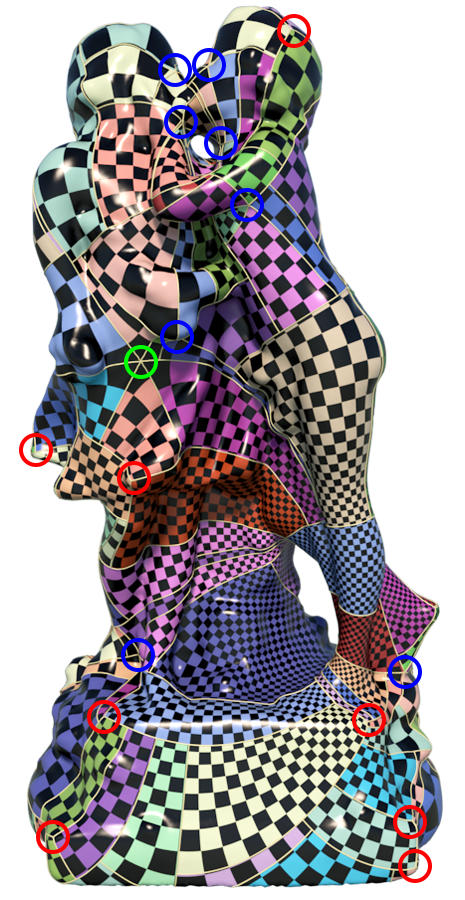}&
        \includegraphics[height=0.60\textwidth]{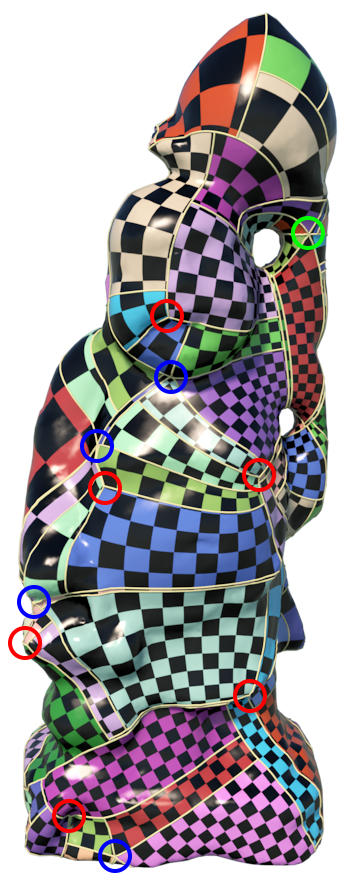}\\
        Witch model& Kiss model& Monk model\\
\end{tabular}
    \caption{Singularities and T-Meshes of the surfaces with complicated geometries.}
    \label{fig:examples_3}
\end{figure}

\begin{figure}
    \centering
\begin{tabular}{ccc}
    \includegraphics[height=0.45\textwidth]{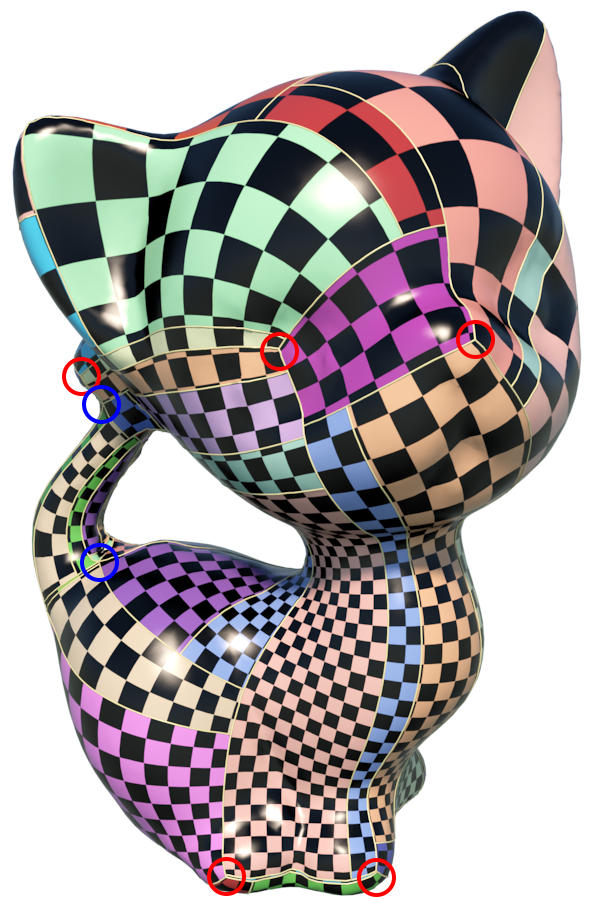}&
        \includegraphics[height=0.45\textwidth]{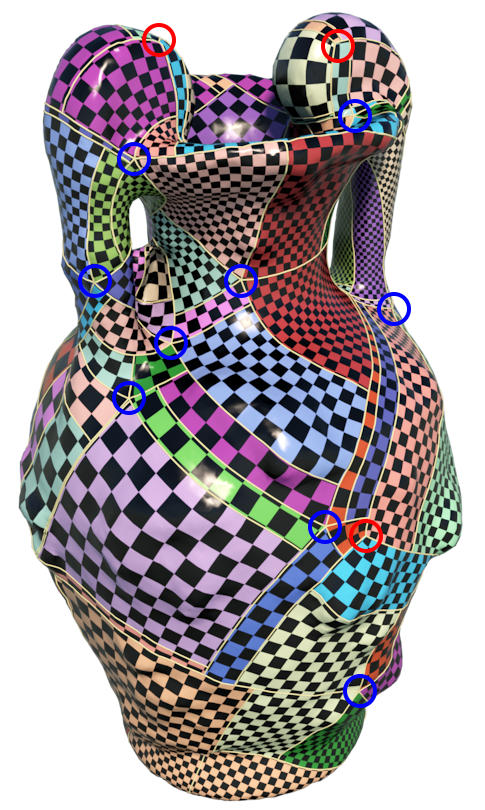}&
    \includegraphics[height=0.45\textwidth]{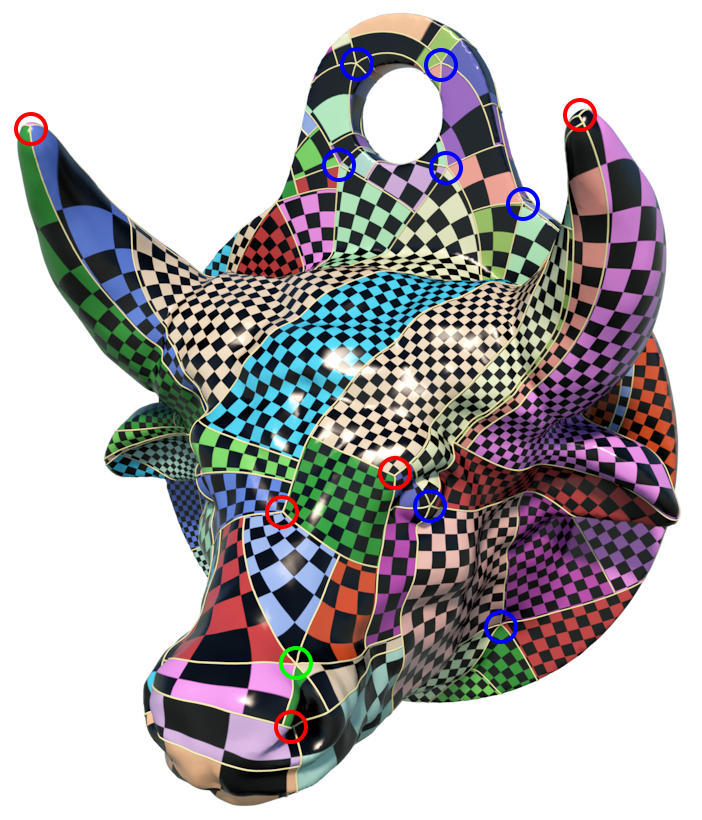}\\
    Kitten model& Amphora model & Bull head\\
\end{tabular}
    \caption{Singularities and T-Meshes of various surfaces.}
    \label{fig:examples_4}
\end{figure}

\begin{figure}
    \centering
    \begin{tabular}{cc}
    \includegraphics[height=0.4\textwidth]{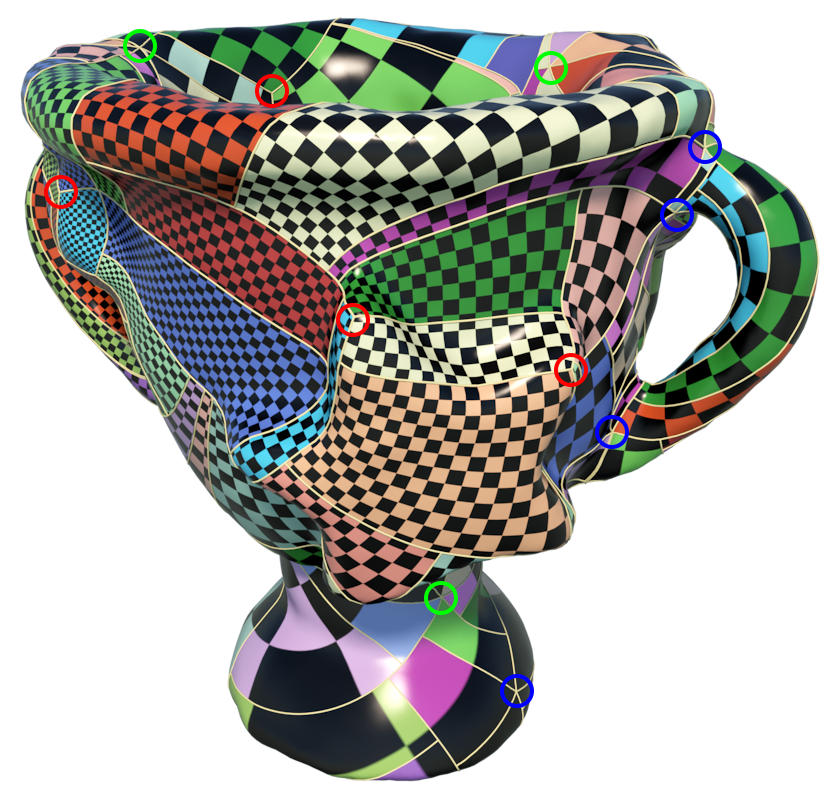}&
    \includegraphics[height=0.35\textwidth]{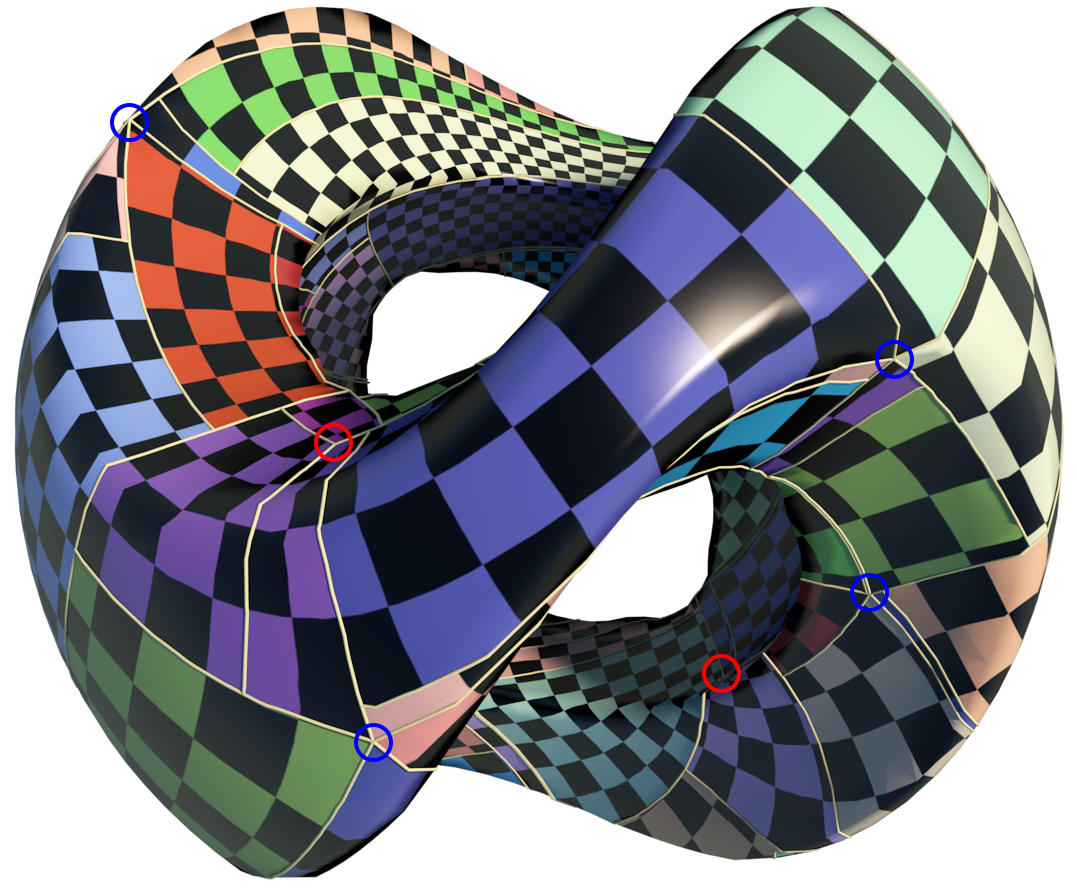}\\
    Star cup & Sculpture model\\
    \end{tabular}
    \caption{Singularities and T-Meshs of high genus surfaces.}
    \label{fig:examples_5}
\end{figure}

\begin{figure}
    \centering
\begin{tabular}{cccc}
    \includegraphics[height=0.45\textwidth]{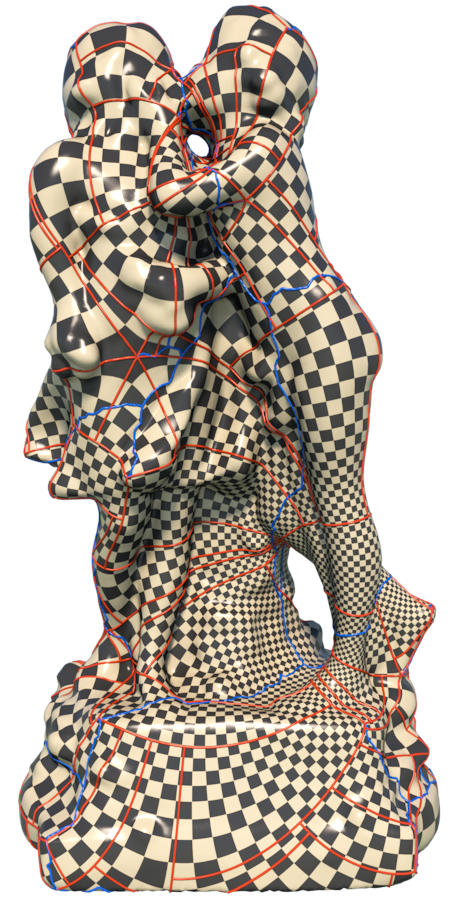}&
    \includegraphics[height=0.45\textwidth]{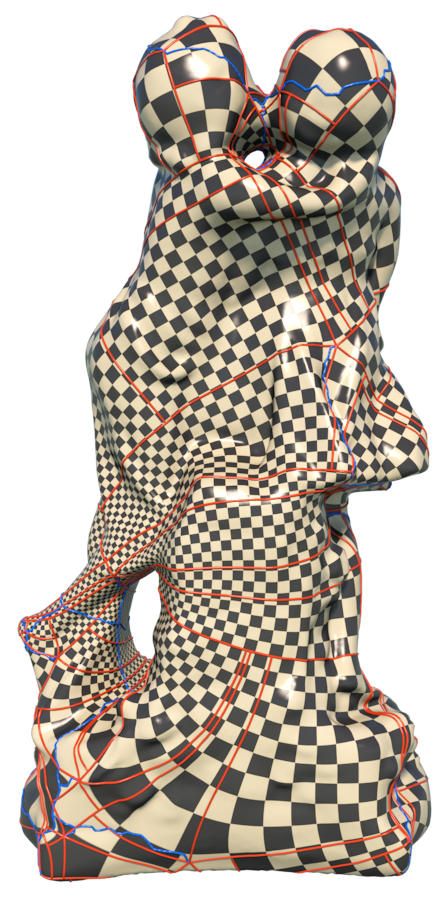}&
    \includegraphics[height=0.45\textwidth]{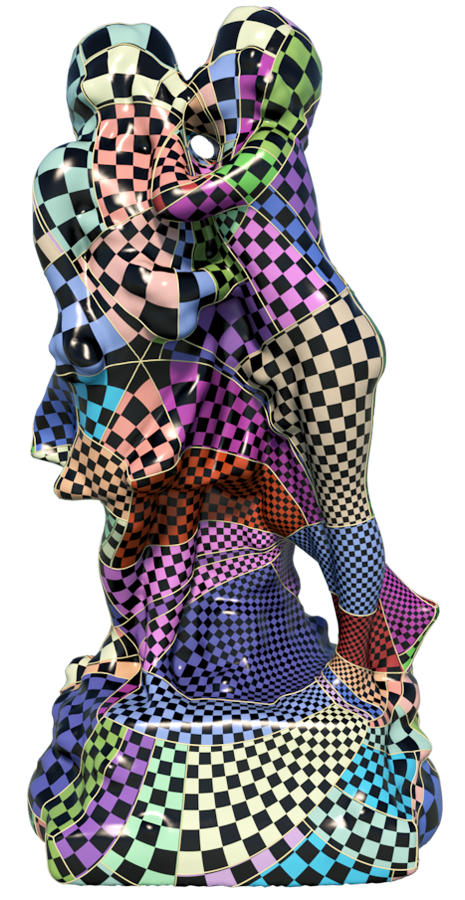}&
    \includegraphics[height=0.45\textwidth]{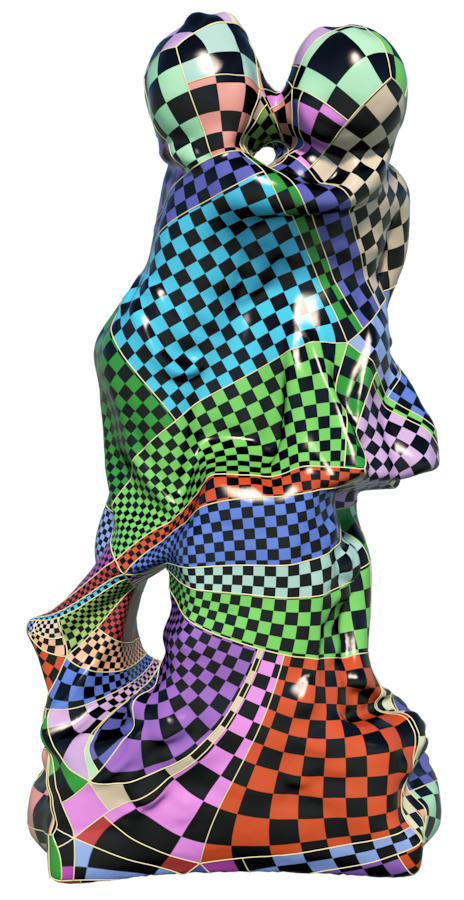}
\end{tabular}
    \caption{The motor-graph and T-mesh of the genus $3$ 2kids surface.}
    \label{fig:2kids}
\end{figure}

\begin{figure}
    \centering
    \begin{tabular}{cc}
    \includegraphics[height=0.25\textwidth]{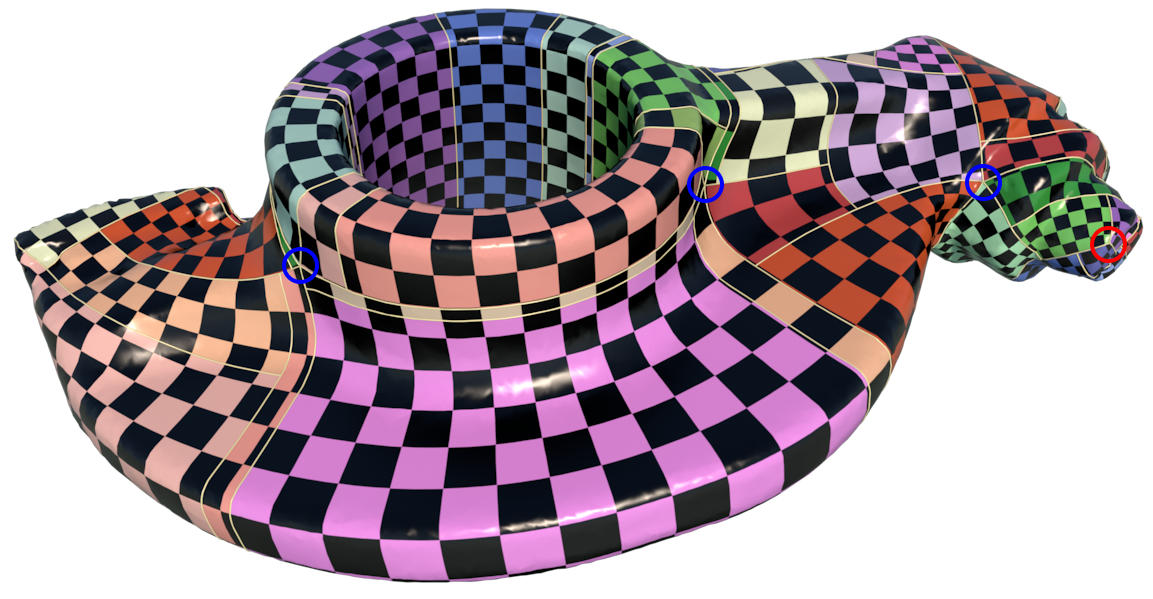}&
    \includegraphics[height=0.25\textwidth]{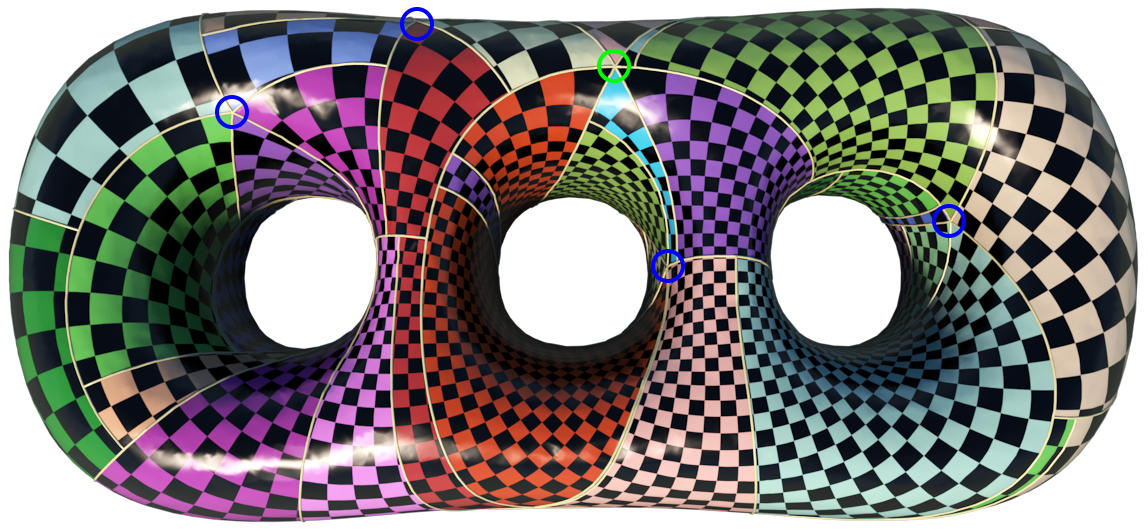}\\
    Rocker arm & 3 holes surface\\
    \end{tabular}
    \caption{Singularities and T-Meshes of high genus surfaces.}
    \label{fig:example_6}
\end{figure}

\begin{center}
\begin{table*}[h!]
\caption{Runing time}
\label{tab:Runing_time}
\resizebox{\textwidth}{35mm}{
	\begin{tabular}{l|c|c|c|c|c|c|c|c|c|c|c}
		\hline\hline
		\multicolumn{1}{c|}{\multirow{2}*{\textbf{Model}}}
        & \multicolumn{4}{|c|}{\textbf{Mesh Information}}
        & \textbf{Holo 1-form}
        & \textbf{Holo zeros}
        & \multicolumn{3}{|c|}{\textbf{Legalization of Singularities} }
        & \textbf{Ricci Flow}
        & \textbf{Iso. Immersion} \\
		\cline{2-12}
		\multicolumn{1}{l|}{~}
        &a
        &b
        &c
        &d
        & \textbf{Time(sec.)}
        & \textbf{Time(sec.)}
        &\textbf{Error Threshold}
        &\textbf{Iterations}
        &\textbf{Time(sec.)}
        &\textbf{Time(sec.)}
        &\textbf{Time(sec.)}  \\
		\hline
        \multicolumn{1}{l|}{\textbf{Kitten}}
        & \textbf{10.2k}
        & \textbf{30.7k}
        & \textbf{20.4k}
        & \textbf{1}
        & \textbf{10.247}
        & ---
        & \textbf{3.0e-4}
        & \textbf{2132}
        & \textbf{0.002}
        & \textbf{0.013}
        & \textbf{0.006} \\
         \multicolumn{1}{l|}{\textbf{Ornament}}
        & \textbf{28.8k}
        & \textbf{86.5k}
        & \textbf{57.7k}
        & \textbf{1}
        & \textbf{47.954}
        & \textbf{---}
        & \textbf{3.0e-4}
        & \textbf{3382}
        & \textbf{0.005}
        & \textbf{6.177}
        & \textbf{0.014} \\
        \multicolumn{1}{l|}{\textbf{Rockerarm}}
        & \textbf{40.2k}
        & \textbf{120.5k}
        & \textbf{80.4k}
        & \textbf{1}
        & \textbf{39.014}
        & \textbf{---}
        & \textbf{3.0e-4}
        & \textbf{2049}
        & \textbf{0.004}
        & \textbf{12.134}
        & \textbf{0.021} \\
        \multicolumn{1}{l|}{\textbf{Dancer}}
        & \textbf{43.0k}
        & \textbf{129.1k}
        & \textbf{86.0k}
        & \textbf{1}
        & \textbf{43.913}
        & \textbf{---}
        & \textbf{3.0e-4}
        & \textbf{6069}
        & \textbf{0.005}
        & \textbf{10.0659}
        & \textbf{0.027} \\
        \multicolumn{1}{l|}{\textbf{Bull}}
        & \textbf{75.8k}
        & \textbf{227.3k}
        & \textbf{151.5k}
        & \textbf{1}
        & \textbf{95.904}
        & \textbf{---}
        & \textbf{3.0e-4}
        & \textbf{2313}
        & \textbf{0.003}
        & \textbf{18.160}
        & \textbf{0.054} \\
        \multicolumn{1}{l|}{\textbf{Sculpt}}
        & \textbf{4.0k}
        & \textbf{12.2k}
        & \textbf{8.0k}
        & \textbf{2}
        & \textbf{4.828}
        & \textbf{0.029}
        & \textbf{1.0e-3}
        & \textbf{37601}
        & \textbf{0.052}
        & \textbf{1.485}
        & \textbf{0.002} \\
        \multicolumn{1}{l|}{\textbf{Starcup}}
        & \textbf{30.0k}
        & \textbf{90.0k}
        & \textbf{60.0k}
        & \textbf{2}
        & \textbf{51.682}
        & \textbf{0.301}
        & \textbf{3.0e-4}
        & \textbf{1167}
        & \textbf{0.005}
        & \textbf{6.654}
        & \textbf{0.013} \\
        \multicolumn{1}{l|}{\textbf{Monk}}
        & \textbf{38.5k}
        & \textbf{115.5k}
        & \textbf{77.0k}
        & \textbf{2}
        & \textbf{86.741}
        & \textbf{4.037}
        & \textbf{3.0e-4}
        & \textbf{17551}
        & \textbf{0.108}
        & \textbf{8.624}
        & \textbf{0.022} \\
        \multicolumn{1}{l|}{\textbf{Hermanubis}}
        & \textbf{39.9k}
        & \textbf{119.8k}
        & \textbf{79.9k}
        & \textbf{2}
        & \textbf{96.122}
        & \textbf{1.441}
        & \textbf{3.0e-4}
        & \textbf{17540}
        & \textbf{0.043}
        & \textbf{10.370}
        & \textbf{0.025} \\
        \multicolumn{1}{l|}{\textbf{Amphora}}
        & \textbf{82.6k}
        & \textbf{246.5k}
        & \textbf{164.3k}
        & \textbf{2}
        & \textbf{174.396}
        & \textbf{0.883}
        & \textbf{3.0e-4}
        & \textbf{5129}
        & \textbf{0.016}
        & \textbf{21.4288}
        & \textbf{0.046} \\
        \multicolumn{1}{l|}{\textbf{Loveme}}
        & \textbf{86.7k}
        & \textbf{260.2k}
        & \textbf{173.5k}
        & \textbf{2}
        & \textbf{191.663}
        & \textbf{1.156}
        & \textbf{3.0e-4}
        & \textbf{5776}
        & \textbf{0.020}
        & \textbf{25.7747}
        & \textbf{0.0.57} \\
        \multicolumn{1}{l|}{\textbf{Buddha}}
        & \textbf{59.4k}
        & \textbf{178.1k}
        & \textbf{118.7k}
        & \textbf{3}
        & \textbf{179.568}
        & \textbf{2.623}
        & \textbf{3.0e-4}
        & \textbf{670539}
        & \textbf{2.021}
        & \textbf{13.117}
        & \textbf{0.026} \\
        \multicolumn{1}{l|}{\textbf{2Kids}}
        & \textbf{61.7k}
        & \textbf{185.2k}
        & \textbf{123.5k}
        & \textbf{3}
        & \textbf{201.353}
        & \textbf{6.274}
        & \textbf{3.0e-4}
        & \textbf{94620}
        & \textbf{0.546}
        & \textbf{14.633}
        & \textbf{0.027} \\
        \multicolumn{1}{l|}{\textbf{3Holes}}
        & \textbf{65.0k}
        & \textbf{195.0k}
        & \textbf{130.0k}
        & \textbf{3}
        & \textbf{218.954}
        & \textbf{0.819}
        & \textbf{1.0e-3}
        & \textbf{343710}
        & \textbf{0.9333}
        & \textbf{16.756}
        & \textbf{0.032} \\
        \multicolumn{1}{l|}{\textbf{Witch}}
        & \textbf{75.0k}
        & \textbf{225.0k}
        & \textbf{150.0k}
        & \textbf{4}
        & \textbf{363.533}
        & \textbf{10.877}
        & \textbf{3.0e-4}
        & \textbf{304729}
        & \textbf{1.033}
        & \textbf{20.343}
        & \textbf{0.051} \\
        \hline\hline
        \multicolumn{1}{c|}{\multirow{2}*{\textbf{Hardware}}}
        & \multicolumn{9}{c|}{\textbf{CPU}}
        & \multicolumn{2}{c}{\textbf{RAM}}\\
        \cline{2-12}
        \multicolumn{1}{c|}{~}
        &\multicolumn{9}{c|}{\textbf{Intel(R) Core(TM) i5-8250U CPU \bm{$@$} 1.60GHz}}
        &\multicolumn{2}{c}{\textbf{16.0GB}}\\
        \hline\hline
	\end{tabular}
	}
\end{table*}
\end{center}

\subsection{Abel-Jacobi and Holonomy condition verification}
\label{sec:holonomy_condition_validation}

\begin{figure}[h!]
    \centering
\begin{tabular}{cccc}
        \includegraphics[height=0.3\textwidth]{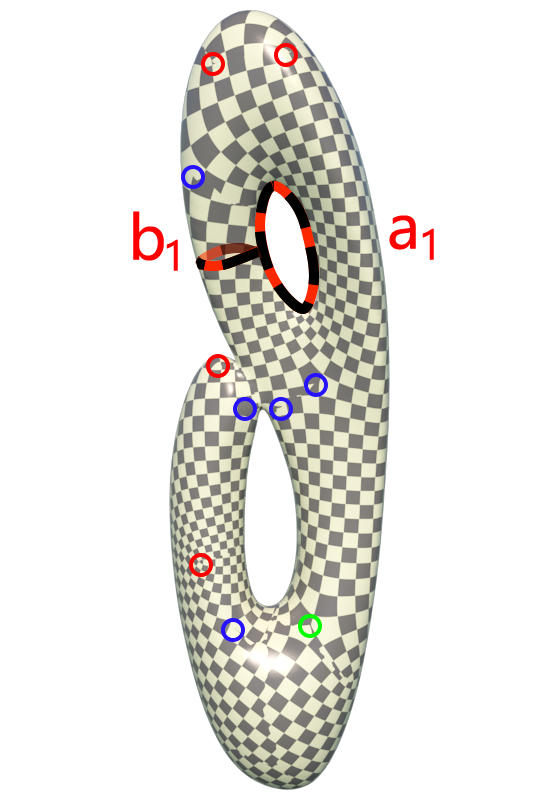}&
    \includegraphics[height=0.3\textwidth]{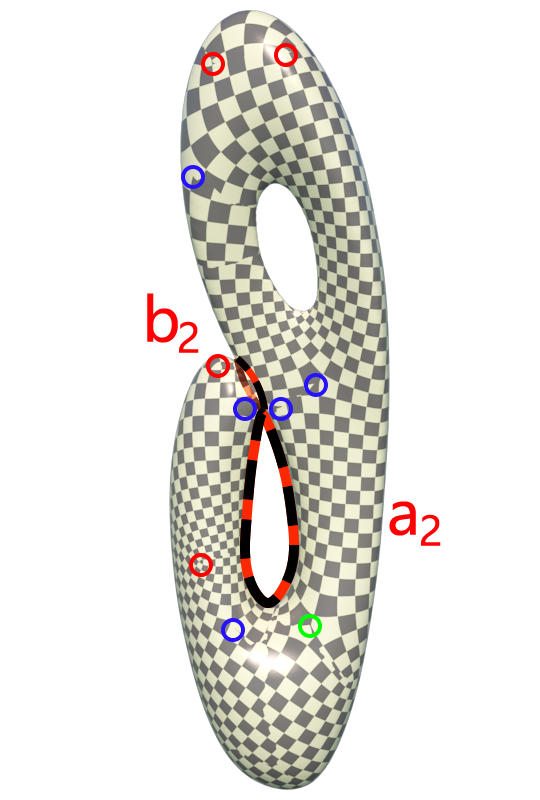}&
    \includegraphics[height=0.3\textwidth]{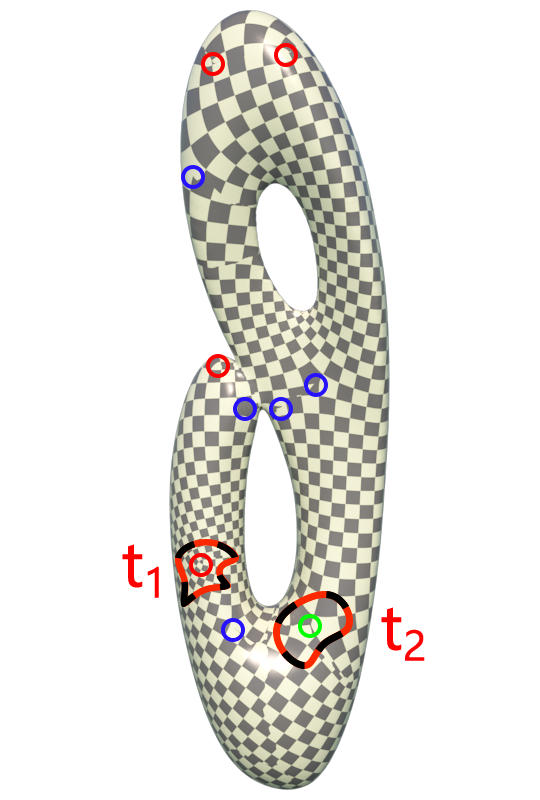}&\includegraphics[height=0.3\textwidth]{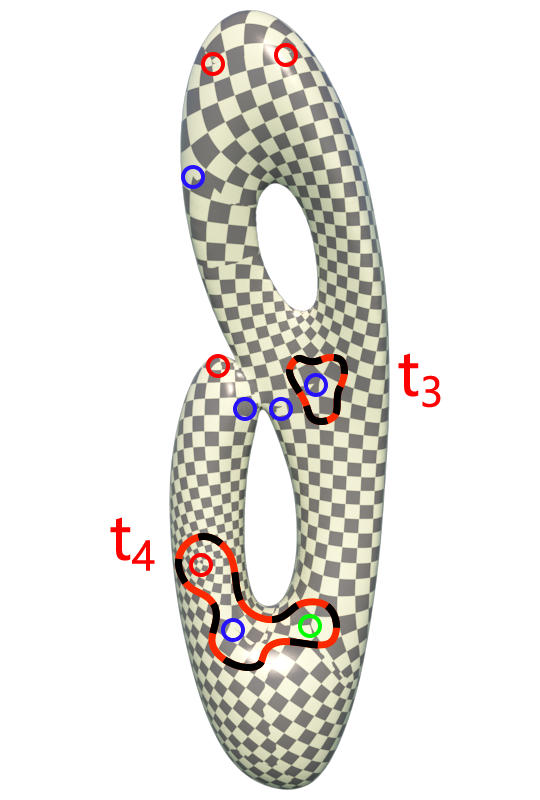}
\end{tabular}
    \caption{The loops on the genus two \textbf{Garniture} model :   $a_1$, $a_2$ are the tunnel loops; $b_1$, $b_2$ are the handle loops; $t_1$, $t_2$, $t_3$ surround index $-1$, $+2$, $+1$ singularities, respectively; $t_4$ encloses three singularities, with index $+1$, $-1$, $-2$ respectively.}
    \label{fig:holonomy_validation}
\end{figure}

\begin{center}
\begin{table}[h!]
\caption{Th holonomy of the loops in Fig.~\ref{fig:holonomy_validation}, rotation components.}
\label{tab:abel_jacobian_mapping_result}
\centering
\small
\renewcommand\arraystretch{0.7}
\centering
\begin{tabular}{p{4cm}<{\centering}|p{1.4cm}<{\centering}|p{1.4cm}<{\centering}|p{1.4cm}<{\centering}|p{1.4cm}<{\centering}}

	\hline\hline
	\textbf{Loops}
    & $\mathbf{a_1}$
    & $\mathbf{b_1}$
    & $\mathbf{a_2}$
    & $\mathbf{b_2}$ \\
    \hline
    \textbf{Rotation degree($^{\circ}$)}
    & 90.31809
    & -0.12269
    & 0.19303
    & 89.81468 \\
    \hline
    \textbf{Loops}
    & $\mathbf{t_1}$
    & $\mathbf{t_2}$
    & $\mathbf{t_3}$
    & $\mathbf{t_4}$\\
	\hline
	\textbf{Rotation degree($^{\circ}$)}
    &  270.00047
    &  540.00136
    &  450.00192
    & 539.99818 \\
    \hline\hline
\end{tabular}
\end{table}
\end{center}
We verify the holonomy conditon for a genus two surface as shown in Fig.~\ref{fig:holonomy_validation}. We compute the tunnel loops $a_1,a_2$ and handle loops $b_1,b_2$, and several loops enclosing different number of singularities. The we compute their holonomies by parallel transportation on the flat metric computed using Ricci flow, the rotation components are reported in the table ~\ref{tab:abel_jacobian_mapping_result}. We can see that all the holonomies are very close to $k90^\circ$, where $k$ is an integer.

Furthermore, for every surface, we compute the image of the singularities under the Abel-Jacobi map, all the results are reported in table ~\ref{tab:abel_jacobian_mapping_result}. We can see that all the images are very close to the zero point in the Jacobian lattice, this shows the singularities satisfy the Abel-Jacobi condition. This demonstrates the accuracy of our proposed algorithm.

\begin{center}
\begin{table*}[h!]
\caption{Abel Jacobian mapping result}
\label{tab:abel_jacobian_mapping_result}
\centering
\small
\renewcommand\arraystretch{0.7}
\centering
\begin{tabular}{p{3cm}<{\centering}|p{8cm}<{\centering}}
	\hline\hline
	\textbf{Model}
    & \textbf{Abel Jacobian Mapping Result} \\
	\hline
    \textbf{KITTEN}
    & $ \begin{pmatrix} \mathbf{ \ 2.13971e-04 \quad  + \quad  i \  * \  7.09315e-05 \ }  \end{pmatrix} $ \\
    ~&~\\
    \textbf{ORNAMENT}
    & $ \begin{pmatrix}\mathbf{ \ -1.09501e-08 \quad  + \quad  i \  * \ 5.73307e-08 \ } \end{pmatrix} $ \\
    \\
    \textbf{ROCKERARM}
   &  $ \begin{pmatrix}\mathbf{ \ -6.05103e-05 \quad  + \quad  i \  * \  6.27266e-06 \ } \end{pmatrix} $ \\
   \\
   \textbf{DANCER}
   & $ \begin{pmatrix}\mathbf{ \ -3.14143e-05 \quad  + \quad  i \  * \  1.57991e-05 \ } \end{pmatrix} $ \\
   \\
    \textbf{BULL}
    & $ \begin{pmatrix}\mathbf{ \ -1.55144e-05 \quad  + \quad  i \  * \ 6.56513e-06 \ } \end{pmatrix} $ \\
    \\
    \multirow{2}*{\textbf{SCULPT}}
    &  \multirow{2}*{$ \begin{pmatrix}\mathbf{ \ -3.72147e-04 \quad  - \quad  i \  * \  9.82485e-04 \ }\\
    \textbf{ \ 8.03122e-04 \quad  + \quad  i \  * \  6.25321e-04 \ }\end{pmatrix} $ } \\
    ~&~\\
    \\
    \multicolumn{1}{c|}{\multirow{2}*{\textbf{STARCUP}}}
    &  \multirow{2}*{ $ \begin{pmatrix}\mathbf{ \ 4.59275e-05 \quad  - \quad  i \  * \  1.27194e-04 \ }\\
    \textbf{ \ 8.14751e-05 \quad  - \quad  i \  * \  2.32289e-04 \ }\end{pmatrix} $ } \\
    ~&~\\
    \\
    \multicolumn{1}{c|}{\multirow{2}*{\textbf{MONK}}}
    &  \multirow{2}*{ $ \begin{pmatrix}\mathbf{ \ -1.37142e-05 \quad  - \quad  i \  * \ 1.84819e-04 \ }\\
    \textbf{ \ 4.70251e-05 \quad  + \quad  i \  * \  1.90921e-04 \ } \end{pmatrix} $ } \\
    ~&~\\
    \\
    \multicolumn{1}{c|}{\multirow{2}*{\textbf{HERMANUBIS}}}
     &  \multirow{2}*{ $ \begin{pmatrix}\mathbf{ \ -1.05753e-04 \quad  - \quad  i \  * \  8.17228e-05 \ }\\
    \textbf{ \ 9.29236e-05 \quad  + \quad  i \  * \  4.96067e-05 \ }\end{pmatrix} $ } \\
    ~&~\\
    \\
    \multicolumn{1}{c|}{\multirow{2}*{\textbf{AMPHORA}}}
     &  \multirow{2}*{ $ \begin{pmatrix}\mathbf{ \ 1.16072e-04 \quad  - \quad  i \  * \  1.37645e-04 \ }\\
    \textbf{ \ 1.32789e-05 \quad  - \quad  i \  * \  1.56983e-04 \ } \end{pmatrix} $ } \\
    ~&~\\
    \\
    \multicolumn{1}{c|}{\multirow{2}*{\textbf{LOVEME}}}
     &  \multirow{2}*{ $ \begin{pmatrix}\mathbf{ \ -9.65795e-05 \quad  + \quad  i \  * \  3.60684e-05 \ }\\
    \textbf{ \ -3.69644e-05 \quad  - \quad  i \  * \  1.48141e-04 \ } \end{pmatrix} $ } \\
    ~&~\\
    \\
    \multicolumn{1}{c|}{\multirow{3}*{\textbf{BUDDHA}}}
    &  \multirow{3}*{ $ \begin{pmatrix}\mathbf{ \ 1.16965e-04 \quad  + \quad  i \  * \  2.90814e-04 \ }\\
    \textbf{ \ -1.28974e-04 \quad  - \quad  i \  * \  7.77251e-06 \ }\\
   \textbf{ \ 1.55074e-04 \quad  - \quad  i \  * \  2.54977e-04 \ }
    \end{pmatrix} $ } \\
    ~&~\\
    ~&~\\
    \\
    \multicolumn{1}{c|}{\multirow{3}*{\textbf{2KIDS}}}
   &  \multirow{3}*{ $ \begin{pmatrix}\mathbf{ \ 2.90402e-04 \quad  - \quad  i \  * \ 2.89651e-04 \ }\\
   \textbf{ \ 2.13554e-04 \quad  - \quad  i \  * \  5.80312e-05 \ }\\
    \textbf{ \ 1.70373e-04 \quad  + \quad  i \  * \  2.77541e-04 \ }
    \end{pmatrix} $ } \\
    ~&~\\
    ~&~\\
    \\
    \multicolumn{1}{c|}{\multirow{3}*{\textbf{3HOLES}}}
   &  \multirow{3}*{ $ \begin{pmatrix}\mathbf{ \ 6.85741e-05 \quad  + \quad  i \  * \  9.32962e-04 \ }\\
   \textbf{ \ 3.55608e-05 \quad  - \quad  i \  * \  8.67721e-04 \ }\\
   \textbf{ \ -1.36089e-05 \quad  + \quad  i \  * \  5.60214e-04 \ }
    \end{pmatrix} $ } \\
    ~&~\\
    ~&~\\
    \\
    \multicolumn{1}{c|}{\multirow{4}*{\textbf{WITCH}}}
    &  \multirow{4}*{ $ \begin{pmatrix}\mathbf{ \ -1.29378e-04 \quad  - \quad  i \  * \  2.40348e-04 \ }\\
  \textbf{ \ -2.75192e-04 \quad  + \quad  i \  * \  1.98399e-04 \ }\\
    \textbf{ \ 2.23835e-04 \quad  + \quad  i \  * \  2.55373e-04 \ }\\
    \textbf{ \ -2.64736e-04 \quad  + \quad  i \  * \  2.39598e-04 \ }
    \end{pmatrix} $ } \\
    ~&~\\
    ~&~\\
    ~&~\\
    \hline\hline
\end{tabular}

\end{table*}
\end{center}

\vspace{-15mm}
\section{Conclusion}
\label{sec:conclusion}
This work proposes a rigorous and practical algorithm for generating meromorphic quartic differentials for the purpose of quad-mesh generation. We give a variational approach to adjust the divisor by an integer programming to satisfy the Abel-Jacobi condition.

Our experimental results demonstrate that the method can handle surfaces with complicated topology and geometry. The algorithm is efficient and accurate. The resuling T-meshes can be used to contruct T-Splines directly.

In the future, we will further explore how to convert the T-meshes to T-Splines, and further optimize the configurations of singularities to improve the quality of the Spline surfaces.

We will also design algorithms for adjusting the conformal structure of the surface to ensure the finiteness of the trajectories of meromorphic differentials and automatic quad-mesh generation.

\section*{Acknowledgment}

The authors thank the encouragements and inspiring discussions with Dr. Tom Hughes and his students Dr. candidate Kendric Shpeherd.

This work is partially supported by NSFC No. 61907005, 61720106005, 61772105 and 61936002.

\bibliographystyle{plain}
\bibliography{references,quad}

\begin{thebibliography}{10}

\bibitem{Alliez2006Periodic}
Pierre Alliez, Bruno L$\acute{e}$vy, Alla Sheffer, and Nicolas Ray.
\newblock Periodic global parameterization.
\newblock {\em Acm Transactions on Graphics}, 25(4):1460--1485, 2006.

\bibitem{Boier2004Parameterization}
Ioana Boier-Martin, Holly Rushmeier, and Jingyi Jin.
\newblock Parameterization of triangle meshes over quadrilateral domains.
\newblock In {\em Acm International Conference Proceeding Series}, pages
  193--203, 2004.

\bibitem{survey:Bommes2013Quad}
David Bommes, Bruno L$\acute{\rm{e}}$vy, Nico Pietroni, Enrico Puppo, Claudio
  Silva, Marco Tarini, and Denis Zorin.
\newblock Quad-mesh generation and processing: A survey.
\newblock {\em Computer Graphics Forum}, 32(6):51--76, 2013.

\bibitem{Carr2006Rectangular}
Nathan~A Carr, Jared Hoberock, Keenan Crane, and John~C Hart.
\newblock Rectangular multi-chart geometry images.
\newblock In {\em Eurographics Symposium on Geometry Processing}, pages
  181--190, 2006.

\bibitem{CMAME_Quad_Mesh_I}
Wei Chen, Xiaopeng Zheng, Jingyao Ke, Na~Lei, Zhongxuan Luo, and Xianfeng Gu.
\newblock Quadrilateral mesh generation i : Metric based method.
\newblock {\em Computer Methods in Applied Mechanics and Engineering (CMAME)},
  accepted.

\bibitem{Dong2006Spectral}
Shen Dong, Peer~Timo Bremer, Michael Garland, Valerio Pascucci, and John~C
  Hart.
\newblock Spectral surface quadrangulation.
\newblock In {\em ACM SIGGRAPH}, pages 1057--1066, 2006.

\bibitem{Gurung2011SQuad}
Topraj Gurung, Daniel Laney, Peter Lindstrom, and Jarek Rossignac.
\newblock Squad: Compact representation for triangle meshes.
\newblock {\em Computer Graphics Forum}, 30(2):355--364, 2011.

\bibitem{He2009A}
Ying He, Hongyu Wang, Chi~Wing Fu, and Hong Qin.
\newblock A divide-and-conquer approach for automatic polycube map
  construction.
\newblock {\em Computers $\&$ Graphics}, 33(3):369--380, 2009.

\bibitem{Huang2008Spectral}
Jin Huang, Muyang Zhang, Jin Ma, Xinguo Liu, Leif Kobbelt, and Hujun Bao.
\newblock Spectral quadrangulation with orientation and alignment control.
\newblock {\em Acm Transactions on Graphics}, 27(5):1--9, 2008.

\bibitem{JFH}
T.~Jiang, X.~Fang, J.~Huang, H.~Bao, Y.~Tong, and M.~Desbrun.
\newblock Frame field generation through metric customization.
\newblock {\em Acm Transactions on Graphics}, 34(4):1--11, 2015.

\bibitem{K2010QuadCover}
Felix K\"alberer, Matthias Nieser, and Konrad Polthier.
\newblock Quadcover -- surface parameterization using branched coverings.
\newblock {\em Computer Graphics Forum}, 26(3):375--384, 2010.

\bibitem{Kowalski2013A}
Nicolas Kowalski, Franck Ledoux, and Pascal Frey.
\newblock A pde based approach to multidomain partitioning and quadrilateral
  meshing.
\newblock In {\em International Meshing Roundtable}, 2013.

\bibitem{L2010Lp}
Bruno L$\acute{e}$vy and Yang Liu.
\newblock Lp centroidal voronoi tessellation and its applications.
\newblock {\em Acm Transactions on Graphics}, 29(4):1--11, 2010.

\bibitem{Li2006Representing}
Wan-Chiu Li, Bruno Vallet, Nicolas Ray, and Bruno L$\acute{e}$vy.
\newblock Representing higher-order singularities in vector fields on piecewise
  linear surfaces.
\newblock {\em IEEE Transactions on Visualization and Computer Graphics},
  12(5):1315--1322, 2006.

\bibitem{Lin2008Automatic}
Juncong Lin, Xiaogang Jin, Zhengwen Fan, and Charlie C.~L Wang.
\newblock Automatic polycube-maps.
\newblock In {\em International Conference on Advances in Geometric Modeling
  and Processing}, pages 3--16, 2008.

\bibitem{Marco2010Practical}
Tarini Marco, Pietroni Nico, Cignoni Paolo, Panozzo Daniele, and Puppo Enrico.
\newblock Practical quad mesh simplification.
\newblock {\em Computer Graphics Forum}, 29(2):407--418, 2010.

\bibitem{Palacios2007Rotational}
Jonathan Palacios and Eugene Zhang.
\newblock Rotational symmetry field design on surfaces.
\newblock {\em Acm Transactions on Graphics}, 26(3):55, 2007.

\bibitem{RS:RPS:2014}
Nicolas Ray and Dmitry Sokolov.
\newblock Robust polylines tracing for n-symmetry direction field on
  triangulated surfaces.
\newblock {\em ACM Transactions on Graphics}, 2014.

\bibitem{Remacle2012Blossom}
J.~F. Remacle, J.~Lambrechts, B.~Seny, E.~Marchandise, A.~Johnen, and
  C.~Geuzainet.
\newblock Blossom-quad: A non-uniform quadrilateral mesh generator using a
  minimum-cost perfect-matching algorithm.
\newblock {\em International Journal for Numerical Methods in Engineering},
  89(9):1102--1119, 2012.

\bibitem{Tong2006Designing}
Y.~Tong, P.~Alliez, D.~Cohen-Steiner, and M.~Desbrun.
\newblock Designing quadrangulations with discrete harmonic forms.
\newblock In {\em Eurographics Symposium on Geometry Processing, Cagliari,
  Sardinia, Italy, June}, pages 201--210, 2006.

\bibitem{Velho20014}
Luiz Velho and Denis Zorin.
\newblock {\em 4-8 Subdivision}.
\newblock Elsevier Science Publishers B. V., 2001.

\bibitem{landau}
Ryan Viertel and Braxton Osting.
\newblock An approach to quad meshing based on harmonic cross-valued maps and
  the ginzburg-landau theory.
\newblock 2017.

\bibitem{Wang2008User}
Hongyu Wang, Miao Jin, Ying He, Xianfeng Gu, and Hong Qin.
\newblock User-controllable polycube map for manifold spline construction.
\newblock In {\em ACM Symposium on Solid and Physical Modeling, Stony Brook,
  New York, Usa, June}, pages 397--404, 2008.

\bibitem{Xia2011Editable}
Jiazhil Xia, Ismael Garcia, Ying He, Shi~Qing Xin, and Gustavo Patow.
\newblock Editable polycube map for gpu-based subdivision surfaces.
\newblock In {\em Symposium on Interactive 3D Graphics and Games}, pages
  151--158, 2011.

\end{thebibliography}

\end{document}